\documentclass{article}

\usepackage{arxiv}

\usepackage[utf8]{inputenc} 
\usepackage[T1]{fontenc}    
\usepackage{hyperref}       
\usepackage{url}            
\usepackage{booktabs}       
\usepackage{amsfonts}       
\usepackage{nicefrac}       
\usepackage{microtype}      
\usepackage{amsmath}
\usepackage{cleveref}       
\usepackage{lipsum}         
\usepackage{graphicx}
\usepackage[
  backend=biber,
  style=nature,
  url=false,
  eprint=false,
  doi=true
]{biblatex}
\usepackage{doi}
\AtEveryBibitem{%
  \clearfield{note}%
  \clearfield{accessed}
}
\title{Physics-constrained generative machine learning-based high-resolution downscaling of Greenland's surface mass balance and surface temperature}

\newif\ifuniqueAffiliation

\ifuniqueAffiliation 
\author{ \href{https://orcid.org/0000-0003-4578-5780}{\includegraphics[scale=0.06]{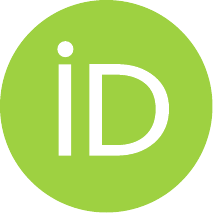}\hspace{1mm}Nils Bochow} \\

}
\else
\usepackage{authblk}

\newbox{\orcid}\sbox{\orcid}{\includegraphics[scale=0.06]{orcid.pdf}} 
\author[1,2]{%
	\href{https://orcid.org/0000-0003-4578-5780}{\usebox{\orcid}\hspace{1mm}Nils~Bochow\thanks{\texttt{contact@nilsbochow.com}}}%
}
\author[2,3]{%
	\href{https://orcid.org/0000-0001-7097-2161}{\usebox{\orcid}\hspace{1mm}Philipp~Hess\thanks{\texttt{philipp.hess@tum.de}}}%
}
\author[4]{%
	\href{https://orcid.org/0000-0003-3519-5293}{\usebox{\orcid}\hspace{1mm}Alexander~Robinson\thanks{\texttt{alexander.robinson@awi.de}}}%
}
\affil[1]{Department of Mathematics and Statistics, Faculty of Science and Technology, UiT The Arctic University of Norway, Tromsø, Norway}
\affil[2]{Potsdam Institute for Climate Impact Research, Potsdam, {Germany}}

\affil[3]{Technical University of Munich, Munich, Germany; School of Engineering \& Design, Earth System Modelling}
\affil[4]{Alfred Wegener Institute, Helmholtz Centre for Polar and Marine Research,  {Potsdam}, {Germany}}
\fi

\renewcommand{\headeright}{Preprint}
\renewcommand{\undertitle}{Preprint}
\renewcommand{\shorttitle}{Physics-Constrained Generative Downscaling for Greenland}

\hypersetup{
pdftitle={Physics-constrained Generative Downscaling of Greenland's SMB},
pdfsubject={physics.geo-ph, cs.AI},
pdfauthor={Nils~Bochow, Philipp~Hess, Alexander~Robinson},
pdfkeywords={Generative modeling, Greenland Ice Sheet, Surface Mass Balance, Downscaling},
}

\begin{document}
\maketitle

\begin{abstract}
Accurate, high‐resolution projections of the Greenland ice sheet’s surface mass balance (SMB) and surface temperature are essential for understanding future sea‐level rise, yet current approaches are either computationally demanding or limited to coarse spatial scales. 
Here, we introduce a novel physics-constrained generative modeling framework based on a consistency model (CM) to downscale low‐resolution SMB and surface temperature fields by a factor of up to 32 (from 160\,km to 5\,km grid spacing) in a few sampling steps. 
The CM is trained on monthly outputs of the regional climate model MARv3.12 and conditioned on ice‐sheet topography and insolation. 
By enforcing a hard conservation constraint during inference, we ensure approximate preservation of SMB and temperature sums on the coarse spatial scale as well as robust generalization to extreme climate states without retraining. 
On the test set, our constrained CM achieves a continued ranked probability score of 6.31\,mmWE for the SMB and 0.1\,K for the surface temperature, outperforming interpolation-based downscaling.
Together with spatial power‐spectral analysis, we demonstrate that the CM faithfully reproduces variability across spatial scales. 
We further apply bias‐corrected outputs of the NorESM2 Earth System Model as inputs to our CM, to demonstrate the potential of our model to directly downscale ESM fields.
Our approach delivers realistic, high‐resolution climate forcing for ice‐sheet simulations with fast inference and can be readily integrated into Earth‐system and ice‐sheet model workflows to improve projections of the future contribution to sea‐level rise from Greenland and potentially other ice sheets and glaciers too.
\end{abstract}


\section{Introduction} 
The Greenland ice sheet (GrIS) is the second largest ice sheet with an ice volume of 7.4\,m sea level equivalent \cite{morlighem_bedmachine_2017}. 
Anthropogenic warming has driven a significant acceleration of Greenland mass loss over recent decades \cite{trusel_nonlinear_2018}. 
Currently, the mass loss of the GrIS is approximately evenly driven by dynamic discharge and a decrease in the surface mass balance (SMB), with ca. 60\% of GrIS ice loss due to the changes at the surface \cite{van_den_broeke_recent_2016, choi_ice_2021}.
However, by the end of this century, the importance of surface processes is projected to increase as marine outlets retreat and dynamic processes will play a smaller role \cite{goelzer_future_2020, choi_ice_2021, payne_future_2021}.
That means that the SMB will be the determining factor of the overall mass balance of the GrIS.
Modeling the GrIS surface mass balance demands fine resolution to capture the steep orographic gradients along the ice-sheet margin that govern precipitation patterns \cite{lucas-picher_very_2012}. 
High‐resolution simulations also resolve the narrow ablation zone and peripheral outlet glaciers, where most of the melt and runoff occur \cite{mottram_surface_2017}. 
These high-resolution SMB fields then feed into ice‐sheet model intercomparisons (e.g., ISMIP6), improving projections of dynamic thinning and sea‐level contribution \cite{goelzer_future_2020}.
Therefore, it is crucial to have realistic and high-resolution projections of the SMB in a changing climate. 

Current methods to calculate the SMB have several disadvantages. 
They are either expensive to run, based on simple parameterisations or have low resolution. 
On short time scales, i.e., centennial scales, regional climate models such as MAR \cite{fettweis_estimating_2013, fettweis_reconstructions_2017} or RACMO \cite{van_dalum_first_2024} are used to simulate the surface mass balance. 
These RCMs are specifically designed to simulate the polar regions. 
However, they are computationally expensive to run and are usually not accounting for a changing ice-sheet topography. 
On longer time scales, the SMB is usually simulated via simple parameterisation schemes such as the positive degree days method (PDD) \cite{aschwanden_contribution_2019, garbe_hysteresis_2020, beckmann_effects_2023} or energy balance models \cite{zeitz_impact_2021, bochow_overshooting_2023, bochow_projections_2024}.
Alternatively, the SMB can be directly derived from ESM runs as a sum of precipitation, melt and runoff in a first approximation \cite{seroussi_ismip6_2020, nowicki_experimental_2020}.
However, the ESM-derived SMB fields are limited to the, often coarse, resolution of the ESM and cannot resolve the steep gradients of the ice sheet, leading to biases in SMB estimates \cite{sellevold_surface_2019}.

Given these computational constraints and resolution trade-offs across centennial and longer scales, computationally fast approaches are required to efficiently produce high-resolution SMB estimates.
Previous approaches to downscaling the SMB of Greenland are often based on statistical downscaling \cite{noel_daily_2016,tedesco_computationally_2023}. 
These approaches leverage the relationship between elevation, SMB and temperature, and have shown promising results in downscaling RCM results to higher resolutions.
However, these statistical methods are inherently limited by their prescribed relationships between melt and elevation, which, for example, do not hold for the Antarctic ice sheet and can therefore not easily be extended to other regions \cite{tedesco_computationally_2023}.
Additionally, these statistical methods cannot easily integrate complementary inputs that could improve the downscaling task \cite{tedesco_computationally_2023}.
Therefore, novel machine-learning techniques, which can be trained to learn complex relationships, including topographic and climatological patterns, from high-resolution simulations and then run orders of magnitude faster are a promising approach.
In contrast to previous statistical approaches, we aim to provide a framework in this work that is able to downscale SMB and $T_s$ fields from other sources than RCMs, that is, deriving high-resolution fields directly from ESM output or from simpler physics-based models. 
Machine learning (ML) methods have gained tremendous attention in climate science in the last years, ranging from reconstruction of climate fields \cite{bochow_reconstructing_2025}, ML-based global circulation and weather models \cite{lam_learning_2023, kochkov_neural_2024},
downscaling \cite{hess_fast_2025} and weather forecasting \cite{bi_accurate_2023, price_probabilistic_2025, watt-meyer_ace2_2025}.
In the glaciological community, ML has, for example, been used to improve ice-flow simulations \cite{jouvet_deep_2022, jouvet_ice-flow_2023}, uncover flow laws of Antarctic ice shelves \cite{wang_deep_2025}, improve upon traditional ice-shelf melt parameterisations \cite{rosier_predicting_2023}, forecast glacier-wide SMB \cite{bolibar_deep_2020, van_der_meer_minimal_2025, steidl_physics-aware_2025} or as emulators for subglacial hydrology models \cite{verjans_accelerating_2024}. 

Generative modeling approaches, and diffusion models in particular, have shown promising results in downscaling of ESM fields \cite{harris_generative_2022, hess_deep_2023, aich_conditional_2024, hess_fast_2025}.
The principle behind downscaling with generative diffusion models is based on so-called denoising, where a network is first trained to remove noise added to a high-resolution sample. Once trained, the network can then be applied for downscaling by iteratively generating realistic small-scale patterns from noise while following the large-scale patterns of the low-resolution ESM.
However, most applications of generative downscaling so far are (i) limited to univariate fields (e.g. precipitation), (ii) struggle with training instabilities or mode collapse \cite{arjovsky_towards_2017}, (iii) require a large amount of sample steps to generate realistic fields \cite{bischoff_unpaired_2024} and/or (iv) lack physical mass conservation \cite{hess_fast_2025}.

\begin{figure}
    \centering
    \includegraphics[width=1\linewidth]{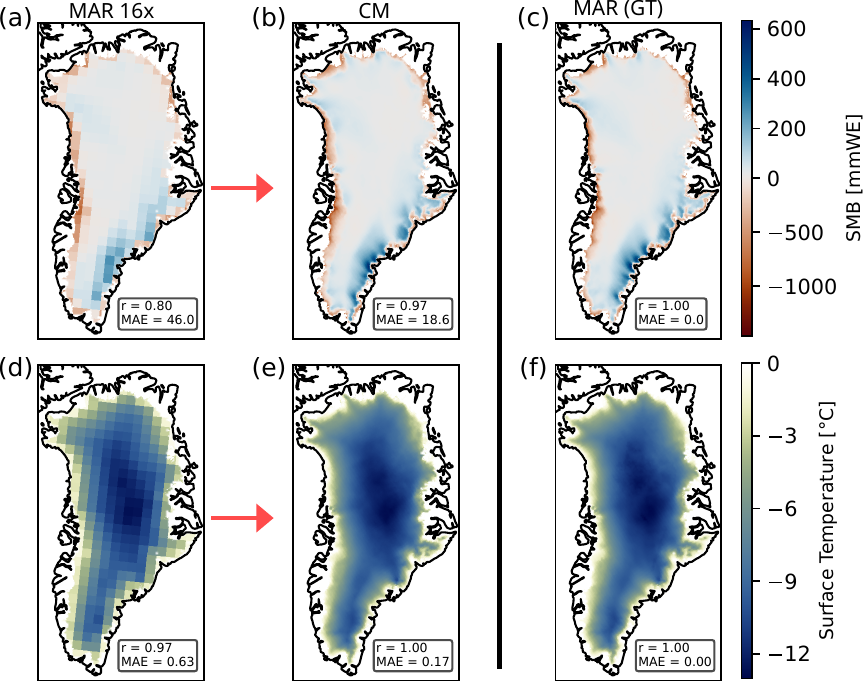}
    \caption{\textbf{Exemplary downscaling of surface mass balance and temperature at surface for random month from test set.} \textbf{(a)} Random month from held-out test set pooled with a factor of 16 (80\,km) resolution. This serves as input to our model. \textbf{(b)} Downscaled SMB field with the CM model using hard constraints to 5\,km resolution. The small scale variability is recovered, especially at the ice-sheet margins. \textbf{(c)} High-resolution ground truth (GT) for this individual month. The GT and the downscaled field are visually indistinguishable. \textbf{(d, e, f)} Same as \textbf{(a, b, c)} but for the temperature at surface. }
    \label{fig:Fig_1a}
\end{figure}

Here, we introduce a generative modeling-based methodology to derive high-resolution surface mass balance and temperature at the surface ($T_s$) fields for the Greenland ice sheet. 
The SMB and $T_s$ are the most important, minimal surface forcing fields needed to drive an ice sheet model.
We use a generative consistency model (CM) \cite{song_consistency_2023, hess_fast_2025} to demonstrate realistic and efficient downscaling of low-resolution SMB and temperature fields by at least a factor of 16 (Fig.~\ref{fig:Fig_1a}, quadrupling the resolution gain of prior work, which was limited to a factor of 4 \cite{hess_fast_2025}. 
In contrast to diffusion models that require dozens or even hundreds of time-steps, a CM directly learns a consistency function that maps a noised version of the target field back to its clean counterpart, allowing for fast generation of samples.
By adding hard constraints during the sampling process, that is, local mass conservation of the SMB and conservation of $T_s$ of the low-resolution field, we are able to generalize to more extreme climate conditions without the need for retraining.
In theory, our model allows the simultaneous downscaling of arbitrarily many variables given high-quality training data.
Compared to previous SMB downscaling approaches, our method does not explicitly prescribe any elevation-SMB relationship.

\begin{figure}
    \centering
    \includegraphics[width=1\linewidth]{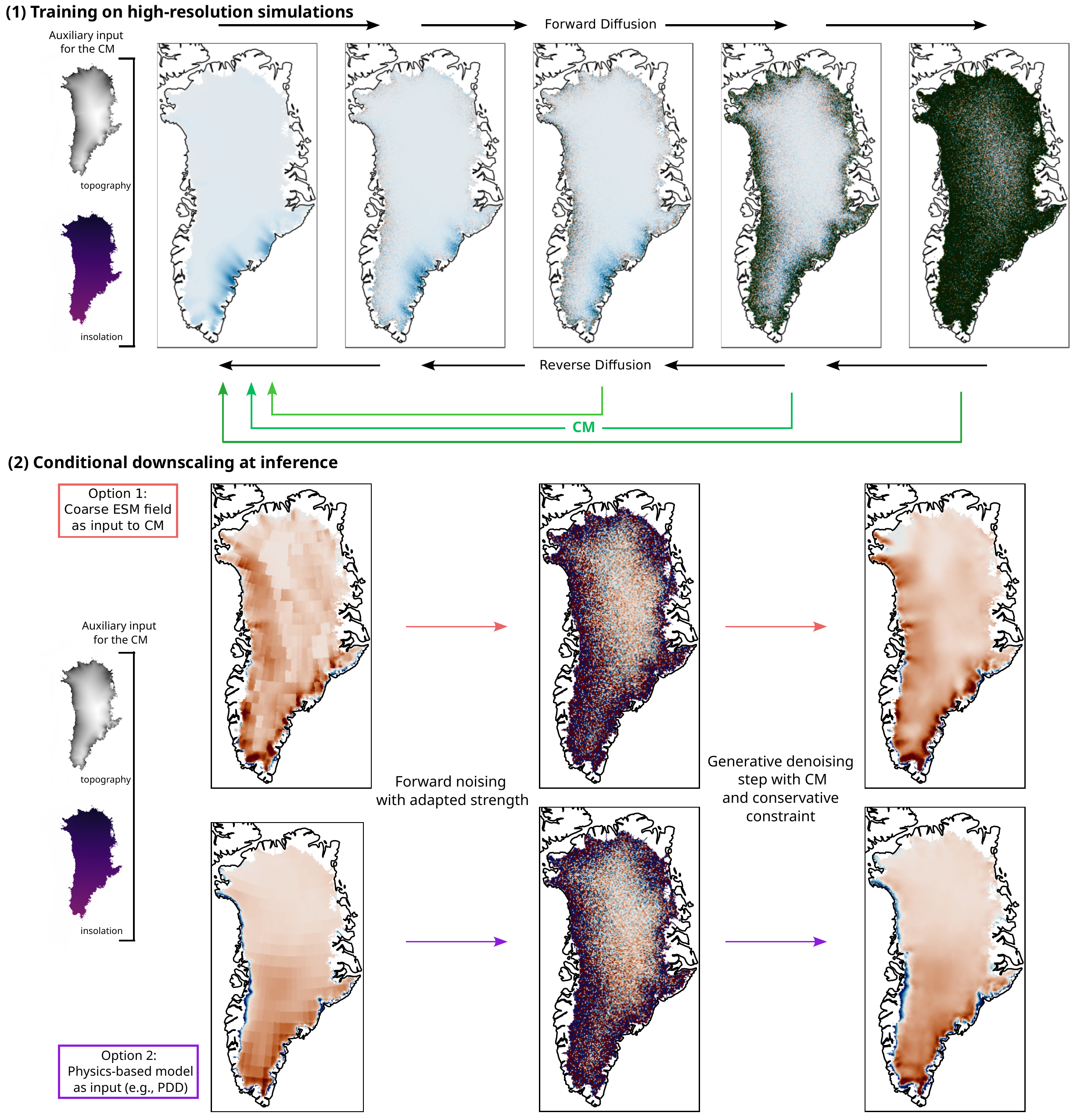}
    \caption{\textbf{Schematic depiction of the consistency model (CM) workflow.} We train the CM on the high-resolution RCM MARv3.12 with topography and insolation as auxiliary input. During inference, the low resolution input fields are noised with an adaptive noising strength and subsequently denoised by the trained CM with conservative constraints. The small-scale variability is recovered by the CM.}
    \label{fig:Fig_2}
\end{figure}

We train our CM on high-resolution simulations of SMB and $T_s$ from the regional climate model MARv3.12 \cite{fettweis_estimating_2013, fettweis_reconstructions_2017}. 
Once trained, the CM can be used to generate high-resolution SMB fields from ESM outputs using either directly the bias corrected SMB fields from the ESM or using a hybrid setup, combining the generative model with a positive degrees days (PDD) approach (Fig.~\ref{fig:Fig_2}). 
We are able to produce realistic SMB and $T_s$ fields with a 5\,km resolution openly available output from ESMs and MAR (Fig.~\ref{fig:Fig_1a}).
Ultimately, our methodology can be used for any climate field and potentially be directly integrated into ESMs.

\section{Results}\label{sec2}
We train and provide two versions of the CM (cf. Methods). 
In the first version, we only downscale the SMB, while in the second version we additionally downscale temperature fields $T_s$. 
In both versions, we include the ice sheet height and the monthly average of the insolation as additional conditioning input. 
Additionally, we implement a hard conservation constraint during inference to conserve the SMB and temperature on a regional level during the downscaling (c.f. Methods).
If not explicitly stated differently, we always show the hard-constrained CM version.
In the following, we concentrate on the SMB downscaling task, since downscaling the temperature is generally easier due to the relative smoothness of the temperature fields.
The model workflow is depicted in Fig.~\ref{fig:Fig_uncond}.
We use monthly output from the RCM MARv3.12 forced by different ESMs over the time period 1950-2100 (~\ref{tab:models_forcing}), in total 23 model runs, i.e., 21432 monthly fields (c.f. Methods).

Here, we evaluate how our model performs on two different downscaling tasks. 
First, we show the downscaling results of the CM using artificially coarsened MAR test set data. This has the advantage that we have paired low- and high-resolution fields for evaluations.
Secondly, we show how the CM can be used to derive high-resolution SMB and $T_s$ fields directly from unpaired low-resolution ESM output.

\subsection{Downscaling low-resolution MAR}
The held-out test set consists of 16 random years between the time period 1950 and 2100.
In total, 2448 months have been held out from the train and validation set. 

To test the ability of the CM to downscale from low to high resolution, we first coarsen the MAR fields by a factor of 16, that is, 16 times lower resolution than the high-resolution fields (80x80\,km). 
This corresponds to a typical spatial resolution as provided by state-of-the-art ESMs.
Additionally, we test our CM approach to downscale spatial fields from a factor 32 lower resolution (160x160\,km).
We downscale these pooled fields with our CM to recover the small-scale variability. 

\begin{figure}
    \centering
    \includegraphics[width=0.8\linewidth]{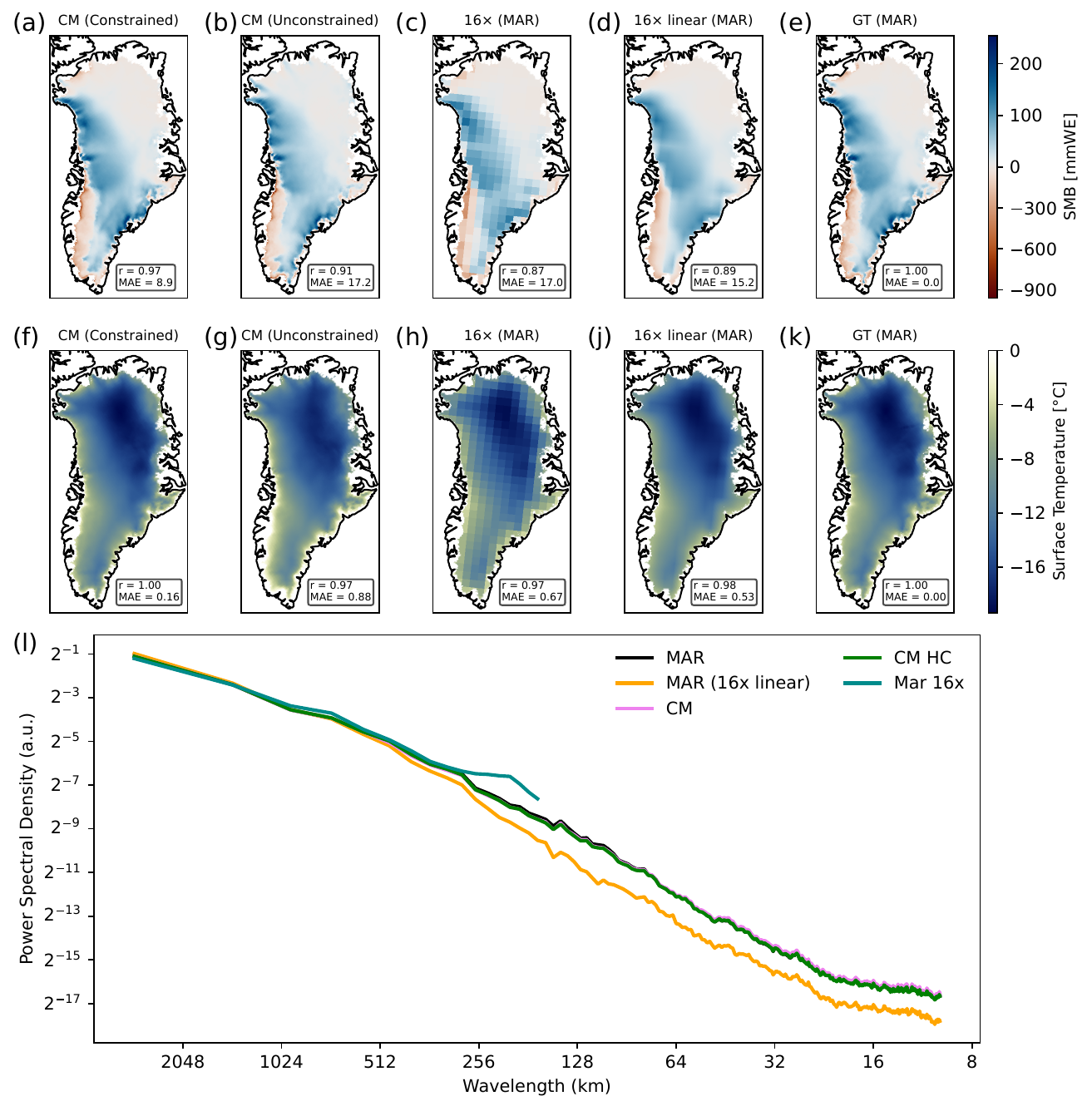}
    \caption{\textbf{Qualitative comparison of downscaled SMB and $T_s$ fields.} \textbf{(a)} Downscaled SMB field of test set with hard constraints during inference for warm month with negative SMB at the margins (blue) with a 5\,km resolution. The mean absolute error (in mmWE) and Pearson correlation with the GT field are denoted. \textbf{(b)} Same as (a) but unconstrained CM. \textbf{(c)} Coarsened MAR field by a factor of 16 (80\,km resolution), which is the input for the CM. The small scale variability is lost. \textbf{(d)} Same as \textbf{b} but linearly interpolated. 
    \textbf{(e)} Ground truth MAR field with 5\,km resolution. The downscaled fields and GT are visually indistinguishable. 
    \textbf{(f-j)} Same as \textbf{a-e} but for the surface temperature. \textbf{(i)}Radially averaged power spectral density of the whole test set for all depicted fields. The coarsened field has no small scale variability (blue). The linear interpolated field shows substantially lower small scale variability compared to the ground truth. The ground truth MAR field and CM fields almost show the same PSD. }
    \label{fig:Fig_4}
\end{figure}

Qualitatively, both the downscaled SMB and $T_s$ fields are visually indistinguishable from the ground truth, while the linearly interpolated fields show a reduced small-scale variability (Fig.~\ref{fig:Fig_1a}\&\ref{fig:Fig_4}).
This is visible in the radially averaged power spectral density (PSD) of the SMB over the whole test set (Fig.~\ref{fig:Fig_4}l). 
The coarse field shows no small-scale variability (blue), while the linearly interpolated field shows a substantial underestimation of the small-scale variability (orange). 
In contrast, the downscaled fields using the CM show a highly accurate representation of small-scale variability compared to the ground truth. 
In particular, the PSD of the hard constrained field (CM HC) is visually indistinguishable from the ground truth.

The mean absolute error (MAE) of one realization of the unconstrained downscaled field compared to the ground truth MAR is $17\,$mmWE/month over the whole test set. 
The mean correlation of the unconstrained CM SMB field is $r=0.84$. 
The hard constrained fields show a substantially lower $\text{MAE}=13.56\,$mmWE/month and higher correlation $r=0.97$ with the ground truth compared to the unconstrained downscaled fields.
Similarly, the hard constrained downscaled $T_s$ field shows the lowest $\text{MAE}=0.16\,$K and correlation $r=1.00$ compared to the ground truth (Fig.~\ref{fig:Fig_4}f). 
For comparison, the linearly interpolated field has a mean absolute error $\text{MAE}=0.53\,$K and correlation of $r=0.98$ (Fig.~\ref{fig:Fig_4}j).

Additionally, we coarsen the MAR fields by a factor 32 (160\,km resolution) and show that the CM, in principle, still can downscale the fields (Fig.~\ref{fig:Fig_32x}).
We show the same exemplary month as in Fig.~\ref{fig:Fig_4} with coarsening factor 16.
While the performance of the CM is still better than simply linearly interpolating, it is worse than for the downscaled field from the coarsened field by factor 16.
The integrated monthly SMB is still mostly conserved with a residual of less than 25\,Gt for almost all months (Fig.~\ref{fig:Fig_32x}b). 
However, due to the very coarse resolution of the input field, some information is lost during the coarsening process.
Again, the error is largest at the margins, especially the southeastern margin (Fig.~\ref{fig:Fig_32x}c,d).
It should be noted that due to the coarse resolution and the overall conservation of the per-cell SMB, the pixel edges are slightly visible in the MAE and the output fields.
This is, for example, visible in southwestern Greenland with negative SMB in Fig.~\ref{fig:Fig_32x}e, where the relatively sharp edge from the coarse field (Fig.~\ref{fig:Fig_32x}f) propagates into the downscaled field. 
This is per construction due to the hard-constraining but could be improved by using the linearly interpolated field as input with the tradeoff of losing information due to the interpolation process.
However, the small spatial scales are still substantially better resolved than in the linearly interpolated field and the error is generally lower (Fig.~\ref{fig:Fig_32x}g).

\subsubsection{Generalizability to extreme climates}
\begin{figure}
    \centering
    \includegraphics[width=1\linewidth]{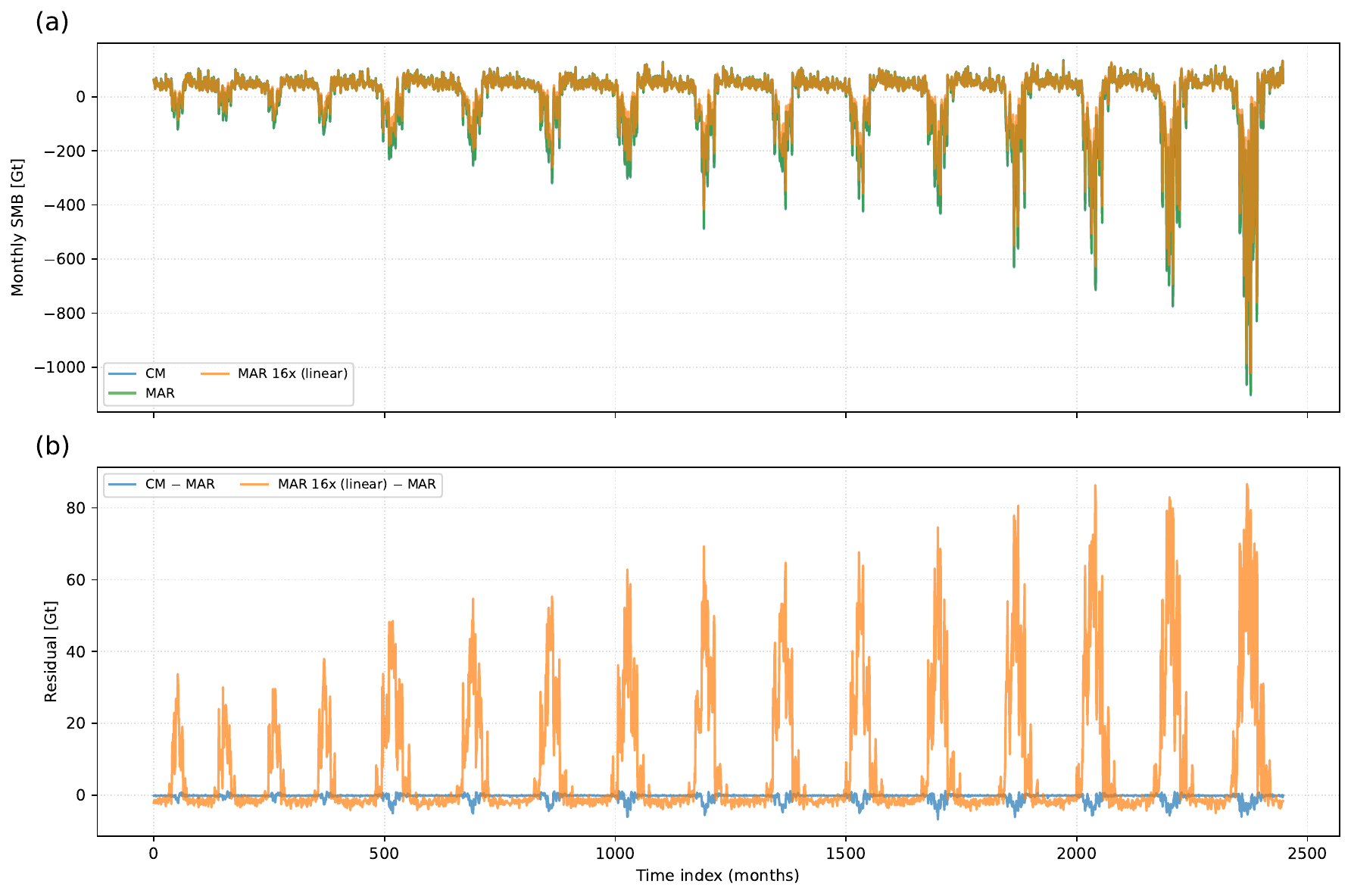}
    \caption{\textbf{Monthly integrated SMB over test set for linearly interpolated SMB and downscaled field.} \textbf{(a)} Monthly integrated SMB over whole test set for MAR, CM and linearly interpolated MAR field from the coarsened MAR field by a factor of 16. \textbf{(b)} Residuals between downscaled field and ground truth MAR and linearly interpolated SMB field and MAR. The linearly interpolated fields overestimates the SMB compared to MAR during the melting seasons. The downscaled fields approximately conserve the SMB even for strong melt. }
    \label{fig:Fig_timeseries}
\end{figure}

Our approach also permits downscaling of fields that are at the far end of the SMB distribution, that is, for high-emission scenarios at the end of this century with extremely negative SMB for most of Greenland.
While the unconstrained model underestimates the melt or overestimates the SMB for most of Greenland (Fig.~\ref{fig:Fig_a1}b), the constrained version is able to realistically downscale the SMB field (Fig.~\ref{fig:Fig_a1}a). 
The MAE for one realization of this specific month is 45\% lower for the constrained CM (MAE=$50.2\,$mmWE) than for a simple linear interpolation (MAE=$90.7\,$mmWE). 
Similarly, the correlation is higher for the constrained CM.
However, the MAE of the unconstrained model (MAE=$335.6\,$mmWE and $r=0.93$) is substantially higher than even for the coarsened field (MAE=$104.9\,$mmWE and $r=0.93$).
For the temperature fields, the linearly interpolated field shows the best metrics (Fig.~\ref{fig:Fig_a1}i).
The MAE of the constrained CM is twice as large (MAE$=0.36\,$K) as for the linearly interpolated field (MAE$=0.18\,$K). 
This is due to artifacts in the downscaled fields outside the ice sheet interior (Fig.~\ref{fig:Fig_a1}f).
However, it should be noted that the MAE is still very small compared to the absolute values of the temperature.
The unconstrained version fails to generate realistic temperature fields (Fig.~\ref{fig:Fig_a1}g).
By hard-constraining the CM, the downscaled fields approximately conserve the SMB even for very negative SMB (Fig.~\ref{fig:Fig_timeseries}).
The maximum absolute error in the monthly integrated SMB of the downscaled field over the whole ice sheet compared to the ground truth is $6.8\,$Gt. 
In contrast, linearly interpolating the coarsened fields leads to a clear overestimation of the SMB for very negative SMB fields (Fig.~\ref{fig:Fig_timeseries}b).
For comparison, the maximum absolute error of the integrated SMB of the linearly interpolated field is $86.6\,$Gt.
For small noising scales ($t<1$) the CM reproduces the hard boundaries of the pooled fields, which can be avoided by linearly interpolating the fields before feeding into the CM.
However, since linear interpolation is not a conserving operation, this leads to a minor loss of information, which ultimately leads to worse quality of the downscaled fields compared to directly downscaling the pooled fields. 

\subsubsection{Determining optimal noising scale}
\begin{figure}
    \centering
    \includegraphics[width=1\linewidth]{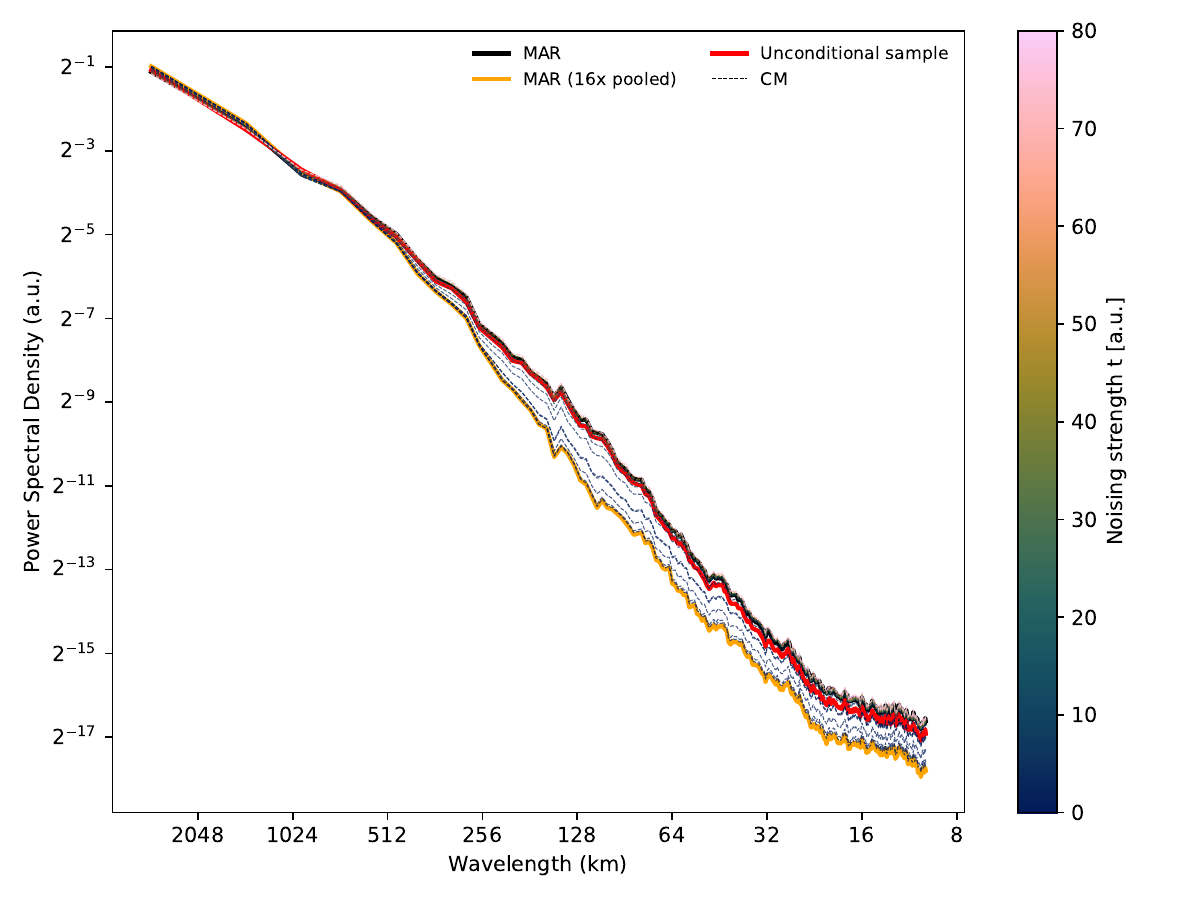}
    \caption{\textbf{Normalized power spectral density (PSD) for different noise scales.} The mean radially averaged normalized PSD for different noising scales is shown. We downscale the test set that was coarsened by a factor of 16 and linearly interpolated using the CM (unconstrained). Additionally, the PSD of an unconditional ensemble is shown (red). The coarsened field (orange) generally underestimates the small scale variability, i.e., the variability below 512\,km. The higher the noising scale, the more the PSD of the downscaled fields resembles the PSD of the original field (black). Larger noise scales correspond to less pairing with the initial fields.}
    \label{fig:Fig_3}
\end{figure}

\begin{figure}
    \centering
    \includegraphics[width=1\linewidth]{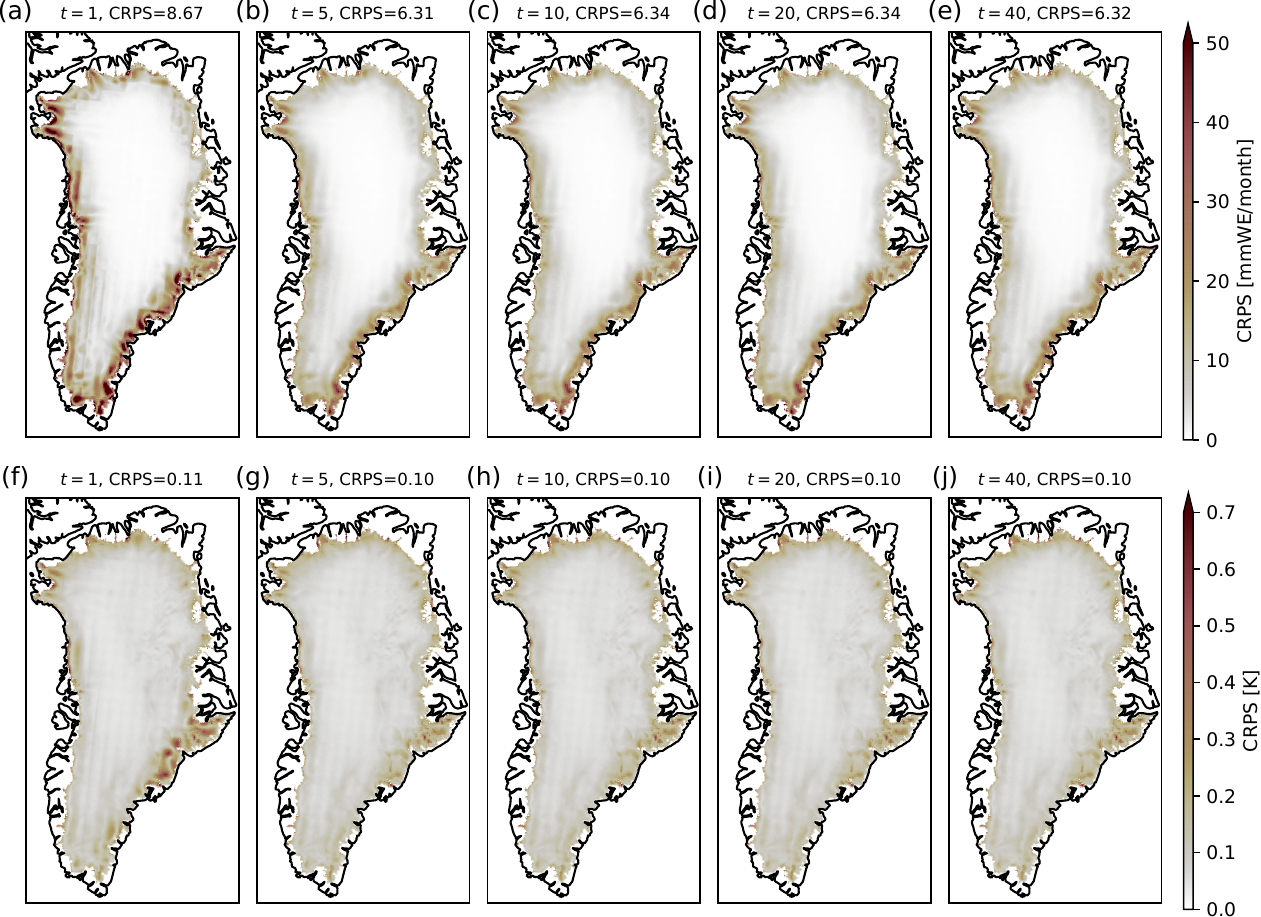}
    \caption{\textbf{Continuous ranked probability scores (CRPS) for different noise scales over the test set for SMB and $T_s$.} \textbf{(a)} Spatial mean CRPS for the noising time $t=1$. For each month of the test set, an ensemble of 50 realizations is generated and compared to the corresponding ground truth. The spatiotemporal mean CRPS is $8.67\,$mmWE/month. The CRPS is largest at the margins of the ice sheet, where the variability is largest. \textbf{(b,c,d,e} Same as \textbf{a} but for different noising times. The noising time $t=5$ leads to the lowest CRPS. \textbf{(f-j)} Same as \textbf{a-e} but for the surface temperature. The CRPS converges to $0.1\,$K for $t>5$.}
    \label{fig:Fig_crps}
\end{figure}

To determine the optimal noising scale, which reasonably recovers the small-scale variability while also providing the smallest errors, we evaluate the power spectral density (PSD) (Fig.~\ref{fig:Fig_3}) and the continuous ranked probability score (CRPS) for different noising scales (Fig.~\ref{fig:Fig_crps}).
While the PSD is a measure of the variability on varying spatial scales, the CRPS is a metric that compares the ground truth to an ensemble of predictions.
It can be seen as a probabilistic generalization of the mean absolute error. 
For this, we generate 50 different realizations for each month of the test set for different noising intensities and calculate the CRPS. 
We find that a noising time of $t=5$ results in the lowest CRPS of $6.31\,$mmWE/month (Fig.~\ref{fig:Fig_crps}b), while also ensuring agreement with the high resolution PSD (Fig.~\ref{fig:Fig_3}).
For noising scales larger than $t=10$, the CRPS converges to $6.34\,$mmWE/month (Fig.~\ref{fig:Fig_crps}e-g).
The CRPS is generally largest at the margins, where the SMB variability is the largest and lowest in the interior with little variability. 
Similarly, the CRPS for the surface temperature is lowest for $t=5$ with CRPS$=0.1\,$K (Fig.~\ref{fig:Fig_crps}g).

\subsection{Downscaling NorESM}
Generally, most ESMs do not directly provide a SMB field. 
However, the SMB can be approximated as the sum of precipitation, evaporation and runoff \cite{seroussi_ismip6_2020, nowicki_experimental_2020}.
Naturally, the CM is able to directly downscale from this first approximation of the SMB.
However, the quality of the SMB field strongly depends on the ESM's ability to correctly simulate these variables.
Most CMIP6 ESM's strongly overestimate the SMB or underestimate the melt in the higher emission scenarios.
Hence, the SMB fields provided by the ESM's are not in accordance with the RCM simulations, even after bias correction (Fig.~\ref{fig:Fig_noresm_qdmwarm}a-d).
The (bias-corrected) temperature, on the other hand, reproduces the overall warming trend shown by MAR reasonably well (Fig.~\ref{fig:Fig_noresm_qdmwarm}e-h). 
We compare the downscaled fields to the MAR fields for the same scenario.
However, it has to be noted that the fields are not paired in the strict sense since MAR is driven by other variables and hence the spatial patterns do not match.
For this downscaling exercise, we use the Norwegian ESM (NorESM2) under the SSP-5.85 forcing scenario, in order to show the general applicability of the approach.
First, we bias correct the surface temperature as well as the SMB fields using Quantile Data Mapping (QDM) with the MAR fields driven by the reanalysis ERA5, using the historical period 1950-2015 as the reference period. 
These fields are then used as input into the constrained CM. 
While the CM is able to realistically downscale the SMB, the fields for the melting season at the end of the century underestimate the melt (Fig.~\ref{fig:Fig_noresm_qdmwarm}c).
This follows from the QDM-corrected SMB fields which show the same behavior and are used as the input (Fig.~\ref{fig:Fig_noresm_qdmwarm}b). 
Per construction, the CM downscaling follows the large-scale patterns and hence cannot correct the low SMB if the input field show a low SMB bias. 
The MAE of the field downscaled by the CM is only slightly lower than that of the QDM-corrected field alone, but it effectively recovers the variability at small spatial scales (Fig.~\ref{fig:Fig_noresm_qdmwarm}\&\ref{fig:Fig_noresm_qdmcold}).
If the bias-corrected fields show very sharp and large transitions between individual grid cells or simply show very unrealistic spatial SMB fields, the conservation of the SMB can lead to wrong values, where they are not expected since the CM is trying to conserve the overall SMB of the locally defined region (superblock).
In other words, the constrained CM cannot simply correct fields that are spatially incoherent.
An example of this behavior is shown in Fig.~\ref{fig:Fig_noresm_qdm_wrong}.
The QDM-corrected field produces an unrealistically narrow accumulation area with positive SMB at the southeastern margin and a strong and abrupt gradient between negative and positive SMB (Fig.~\ref{fig:Fig_noresm_qdm_wrong}b).
While the CM downscales to a realistic SMB at the southeastern margin, the CM then is forced to compensate with a negative SMB west of the accumulation area due to the overall conservation of the SMB (Fig.~\ref{fig:Fig_noresm_qdm_wrong}c).
To avoid this behavior, the input field can be linearly interpolated beforehand or the superblock size has to be tuned accordingly. 

The original temperature fields are substantially too cold (Fig.~\ref{fig:Fig_noresm_qdmwarm}e\&\ref{fig:Fig_noresm_qdmcold}e), while the bias corrected fields show a similar temperature range compared to the MAR fields (Fig.~\ref{fig:Fig_noresm_qdmwarm}f\&\ref{fig:Fig_noresm_qdmcold}f).
This shows the importance of bias-correcting ESM fields for any further analysis.
The CM fields show the smallest MAE with $0.68\,$K for a random warm month and $0.72\,$K for a random cold month.
Over the whole time period, the MAE of the downscaled SMB fields is $38.9\,$mmWE and for the temperature $1.17\,$K, compared to the MAR fields forced by NorESM2. 
The largest differences are at the southeastern and western margins (Fig.~\ref{fig:Fig_noresm_qdm_mae}).

Since the SMB simulated by the NorESM underestimates the melt at the end of this century, we show an additional approach using a simple SMB parameterisation. 
Specifically, we show how our approach can be used in a \textit{hybrid} manner by (i) using a physically-motivated model to get a low-resolution estimate of the SMB and (ii) run the CM to get a high-resolution field from this estimate.
We run a simple offline PDD model with the bias corrected NorESM2 (SSP-5.85) temperature field and the precipitation field as input to get a very simple estimate of the SMB.
For simplicity, we do not bias correct the precipitation field before calculating the SMB, which might improve the SMB further.

We find a MAE of $33.3\,$mmWE over the whole time period and the MAE of the temperature remains $1.17\,$K (Fig.~\ref{fig:Fig_noresm_pdd_mae}a,b).
For comparison, the MAE of the PDD-derived SMB field is slightly higher with $34.6\,$mmWE and $1.2\,$K for the temperature (Fig.~\ref{fig:Fig_noresm_pdd_mae}c,d).
Again, the margins show the largest error and the CM is able to recover the small-scale variability. 
While the PDD approach does not substantially improve the performance compared to the QDM approach, it shows that the CM can be used in a hybrid manner, i.e., using it in combination with a physics-based model.

\section{Discussion}\label{sec13}
Our approach is built as a conditional generative model, not a classical regression, that learns to sample from a distribution of realistic high‑resolution SMB fields given coarse inputs rather than to deterministically predict the SMB from external input such as the temperature or precipitation.
However, similar to a regression, our recovered SMB and $T_s$ fields strongly depend on the quality of the low-resolution input fields, that is, the ability of the ESM or physics-based model to accurately simulate the SMB and $T_s$ at the coarse scale.
We tried including the temperature as additional input to the CM during training (i.e., predictor) for the SMB (not shown). 
While we were able to generate realistic samples, the warming trend and interannual variations were underestimated in future emission scenarios.

A more promising alternative approach would be to separate the regression, i.e. predicting SMB from ESM variables, and downscaling tasks \cite{aich_diffusion_2025}. 
For example, a U-net could be trained to learn a regression from several input variables to give a first low-resolution estimate of the SMB with subsequent downscaling using the CM.
Depending on the chosen regressor variables such an approach would allow a coupling with an external ice sheet model and therefore a first prototype of a hybrid SMB-ice sheet model. 
The CM model we present here is therefore a first step in this direction.

While our approach to constraining the model via mass conservation generally assumes some confidence in the overall low-resolution SMB fields, it allows generalization to more extreme or even out-of-sample SMB fields without the need of retraining the CM.  
This could be used to downscale regional climate models when running on a high-resolution grid is too expensive or time consuming.
The subsequent downscaling step with the CM ensures conservation of the overall SMB fields and trend while providing realistic high-resolution fields at the same time. 
However, we also find that the CM reaches its limit beyond a downscaling factor of 32. 
Alternatively, our model can be used to generate unconditional samples that represent constant present-day forcing with the learned natural variability. 
This can be used for, e.g., spinup simulations, when a constant climate with variability is desirable.

Due to the nature of data-driven approaches, our results are strongly dependent on the quality of the training data. 
Here, we only use the regional climate model MAR, however, additional regional climate models such as RACMO \cite{van_dalum_first_2024} are available. 
It has been shown, that these show variations by a factor of two in future projections of the SMB \cite{glaude_factor_2024}.
Naturally, our downscaled fields are therefore biased towards the MAR projections. 
Nonetheless, our constraining approach overcomes this problem to some extent by enforcing approximate conservation of the regional SMB.
The hard constraining is especially useful in the context of embedding downscaling approaches within existing process-based models such as ESMs.
In ESMs, conservation of, for example, water, mass and energy is important to enforce physical consistency between the individual components.
Our constraining approach ensures conservation of these variables, while allowing higher resolution.
It should be noted, that our constraining approach is relatively simple and can be improved by, for example, incorporating the constraint layer directly into the network architecture, using a different constraint approach (e.g., multiplicative constraints) \cite{harder_hard-constrained_2023} or by simply improving our region-growing algorithm, which might fail for some special cases. 

Unlike precipitation, which is defined (even if zero) everywhere and can be treated as a spatially continuous field, surface mass balance is strictly confined to the ice‐sheet mask.
Downscaling SMB, therefore, must handle sharp domain boundaries at the ice edge, where values abruptly drop to zero, rather than the smooth transitions typical of precipitation fields
This leads to several technical challenges on how to treat the outside of the ice sheet in spatially-aware generative models. 
We decided to simply treat the SMB outside the ice sheet as zero.
In the case of a non-evolving topography, as given by the training data, this is the most natural choice. 
However, this also means that our model is currently restricted to ice sheet geometries that do not extend beyond present-day extent due to the nature of the training data.
With increasing availability of high-resolution RCM simulations with evolving topography this problem might be solved \cite{delhasse_coupling_2024}.
Similarly, the resolution of the output field is mostly limited by the training data.
Training data with even higher resolution would also allow to generate even higher resolution downscaled fields.
However, for very high resolutions, the training might be limited by the available GPU memory.

\section{Conclusion}
This study introduces a physics-constrained generative machine learning approach, based on consistency models to downscale the spatial SMB and $T_s$ fields of the GrIS.
We train our model on the regional climate model MARv3.12 and show that our approach is able to downscale the SMB and $T_s$ fields by a factor of up to 32, from 160\,km resolution to 5\,km resolution in an efficient and realistic manner. 
The resulting fields are visually indistinguishable from the ground truth and the CM is able to successfully recover the small-scale variability of the SMB and $T_s$ fields.
Furthermore, we show how our model can be used to directly derive high-resolution SMB and $T_s$ fields from the ESM by combining the CM with either (i) a bias-correction method (QDM) or (ii) a physics-based method (PDD).

Our approach is a first step towards a more general machine-learning driven SMB model. 
An application of our approach to Antarctica is straightforward but likely needs retraining and corresponding high-quality training data.
Alternatively, our approach could be integrated into ESMs either to downscale SMB fields provided by the land component or to directly downscale precipitation fields.
Usually, the resolution of the precipitation fields generated by the atmospheric component is several times lower than the native resolution of the ice sheet model. 
Hence, the precipitation does not resolve the topography of the ice sheets at all, which in turn leads to unrealistic spatial precipitation distribution and ultimately SMB.
Using the CM with hard-constraints to downscale the precipitation fields during runtime could solve this problem. 
Overall, generative modeling is a promising approach to downscale arbitrary climate fields in computationally fast manner.

\section{Methods}
\subsection{Consistency model}
We use a consistency model architecture \cite{song_consistency_2023}, recently introduced for downscaling \cite{hess_fast_2025} to derive high-resolution SMB and temperature fields. 
CMs are inspired by iterative diffusion models \cite{ho_denoising_2020, song_score-based_2021}, but offer several advantages.
The idea behind CM is to learn a direct mapping from noisy inputs to clean data such that any point along the same noise trajectory (for the same underlying data) maps to the same clean state \cite{song_consistency_2023}. 
In contrast to previously introduced diffusion models, which need many evaluation steps (usually in the range of 10-2000) to generate realistic samples, CM can therefore generate realistic samples in only one step.
Consistency models learn a consistency function $ f(x(t), t, y)=x(t_{\text{min}})$, which is self consistent:
\begin{align}
    f({\bf{x}}(t),t, {\bf{y}})=f({\bf{x}}({t}^{{\prime}}),{t}^{{\prime} },  {\bf{y}})\forall t,{t}^{{\prime} }\in [{t}_{\min },{t}_{\max }],
\end{align}
with the time-independent conditioning fields ${\bf{y}}$, the input fields $\bf{x}(t)$, i.e., the $T_s$ and SMB and the time $t$.
That is, the model’s output is invariant to the noise level.
Following Song et al. (2023) and Hess et al. \cite{song_consistency_2023, hess_fast_2025}, we set $t_{\text{min}}=0.002$ and $t_{\text{max}}=80$.
We impose the boundary condition $f(x(t_{\text{min}}), t_{\text{min}})=x(t_{\text{min}})$ for $t=t_{\text{min}}$.
This is parameterised in the model via skip connections 
\begin{align}
    f({\bf{x}},t, {\bf{y}};{{\theta }})={c}_{{\rm{skip}}}(t){\bf{x}}+{c}_{{\rm{out}}}(t)F({\bf{x}},t,{\bf{y}};{{\theta }}),
\end{align}
where $F(\cdot)$ is a U-Net with parameter $\theta$.

The parameters $ {c}_{{\rm{skip}}}$ and ${c}_{{\rm{out}}}$ are defined as 
\begin{align}
    {c}_{{\rm{skip}}}=\frac{{\sigma }_{{\rm{data}}}^{2}}{({(t-{t}_{\min })}^{2}+{\sigma }_{{\rm{data}}}^{2})},\qquad {c}_{{\rm{out}}}(t)=\frac{{\sigma }_{{\rm{data}}}t}{\sqrt{{t}^{2}+{\sigma }_{\rm{data}^{2}}}}.
\end{align}

The consistency training objective is given by \cite{song_consistency_2023}
\begin{align}
    {\mathcal{L}}\left({{\theta }},\bar{{{\theta }}}\right)={{\mathbb{E}}}_{{\bf{x}},n,{t}_{n}}\left[d\left(\,f({\bf{x}}+{t}_{n+1}{\bf{z}},{t}_{n+1}, {\bf{y}};{{\theta }}),f\left({\bf{x}}+{t}_{n}{\bf{z}},{t}_{n}, {\bf{y}};\bar{{{\theta }}}\right)\right)\right].
\end{align}
The distance measure $d(\cdot,\cdot)$ is given by $d({\bf{x}},{\bf{y}})={\rm{LPIPS}}({\bf{x}},{\bf{y}})+| | {\bf{x}}-{\bf{y}}| {| }_{1}$ \cite{song_consistency_2023, hess_fast_2025}, where LPIPS is the Learned Perceptual Image Patch Similarity \cite{zhang_unreasonable_2018}.

The training objective is minimized by stochastic gradient descent (Adam optimizer) on the model parameters $\theta$, while updating 
$\bar{{{\theta }}}$ with an exponential moving average (EMA): 
\begin{align}
    \bar{{{\theta }}}={\rm{stopgrad}}[w(k)\bar{{{\theta }}}+(1-w(k){{\theta }})].
\end{align}
The decay schedule $w(k)$ is given by 
\begin{align}
    w(k)=\exp \left(\frac{{s}_{0}\log \left({w}_{0}\right)}{N(k)}\right),
\end{align}
with $w_0=0.9$.
The EMA has been found to greatly stabilize training and improve final performance of the CM \cite{song_consistency_2023}.

While, CMs can be distilled from diffusion models, we train the model from scratch. 
We use the same U-Net backbone as in Hess et al. (2025) \cite{hess_fast_2025}. 
We train for 200 epochs with a batch size of 4.
Training on a single NVIDIA H100 Tensor Core GPU takes approximately 5 days. 

While, CMs can generate samples in only one step, we use multi-step sampling (max. 10 steps) to enhance sample quality and avoid artifacts \cite{song_consistency_2023}. 
Starting from an initial noisy input, we progressively add scaled Gaussian noise at discrete time steps and denoise the field using the CM.
One evaluation, i.e. downscaling one monthly field, for $t=5$ with hard constraining takes ca. 1.6\,s on a NVIDIA H100 Tensor Core GPU.

\subsection{Bias Correction}
We use Quantile Data Mapping (QDM) for bias correcting the ESM fields when used as input for the PDD or CM using the Python library xclim \cite{bourgault_xclim_2023}, which follows the algorithm introduced by Cannon et al. (2015) \cite{cannon_bias_2015}.
The idea of QDM is to correct systematic biases in modeled time series compared to a reference period of observed data or ground truth (here the RCM MAR) while conserving  model-projected relative changes in quantiles \cite{cannon_bias_2015}. 
It is applied for each spatial grid cell and variable individually.
The QDM algorithm uses two steps. 
First, the future model output is detrended by quantile and bias corrected to the reference data by quantile mapping. 
Second, the absolute or relative change in quantiles is superimposed on the bias-corrected model output. 
The time-dependent cumulative density function (CDF) $\tau_{m,p}(t) = F_{t}(x_{m,p}(t))$ of the modeled projected time series $x_{m,p}$ over a time window is calculated (e.g., the future period). 
The relative change in quantiles between the modeled reference period $\text{ref}$ and a future time $t$ is given by 
\begin{align}
    \Delta_m(t) = \frac{F_t^{-1}(\tau_{m,p}(t))}{F^{(-1)}(\tau_{m,\text{ref}}(t))} = \frac{x_{m,p}(t)}{F^{(-1)}(\tau_{m,\text{ref}}(t))}
\end{align}
with the inverse CDF $F^{-1}$.
The modeled $\tau_{m,p}(t)$ quantile is then bias corrected by applying the inverse CDF derived from the observed values over the reference period 
\begin{align}
    \hat x_{o:m, \text{ref}:p}(t) = F^{(-1)}_{o, \text{ref}}(\tau_{m,p}(t)).
\end{align}
Finally, the bias-corrected modeled projected time series at time $\hat x_{m,p}(t)$ is given by 
\begin{align}
    \hat x_{m,p} = \hat x_{o:m, \text{ref}:p}(t) + \Delta_m(t).
\end{align}
We use 100 quantiles and do a monthly-based bias correction.

\subsection{Conservation Constraints}
To allow the CM to downscale fields that are far out-of-sample, we implement the option to enforce a hard conservation constraint during the inference \cite{harder_hard-constrained_2023}.
This does not require a retraining of the CM. 

The initial conditions are pooled using adaptive pooling with a factor that should correspond to the resolution ratio between the low and high resolution fields. 
This gives a field with so-called superblocks, i.e., pooled pixels.
The first denoising step is not changed and gives a first guess of the downscaled field. 
After the first denoising step, we add the residual between each superblock and the corresponding high-resolution pixels. 
That is, to each intermediate output pixel during the multistep sampling $\tilde y_i$ the residual between the superblock $x$ and the mean of the corresponding pixels $\frac{1}{n}\sum_{i=1}^n \tilde y_i$ is added:
\begin{align}
    y_j \;=\; \tilde y_j + x \;-\; \frac{1}{n}\sum_{i=1}^n \tilde y_i.
\end{align}
The residuals are corrected to account for the number of land pixels in each superblocks. 
That is, the residuals are only distributed onto the valid GrIS pixels and not into the ocean.
In the last denoising step, we do not enforce this constraint. 
To avoid artifacts in the downscaled field due to the pooling step, we use the \textit{area} interpolation and Gaussian filter on the residual field before adding the residual field in each step. 

To avoid cases where all residuals of a superblock are distributed to only a few pixels of the GrIS, which would lead to unrealistic SMB values at the GrIS margins, we implement a simple greedy region‐growing algorithm. 
Each superblock cell that has less than a chosen number of corresponding land pixels in the high-resolution field (we choose 16$^2$ pixels) annexes its largest neighboring blocks until it exceeds the chosen threshold.
This continues until all superblocks are assigned a region with at least 16$^2$ pixels. 
We find that 16$^2$-32$^2$ pixels is a reasonable choice ensuring realistic fields and avoiding artifacts in the downscaled fields.

Since the conservation of the SMB should happen in physical units, we transform the correction field and intermediate fields back to physical units before we apply the constraint.
Subsequently, we transform the fields back to the transformed units before running the next step of the CM. 

\subsection{PDD Model}
We use a simple positive degree day (PDD) model based on the PyPDD implementation  \cite{seguinot_pypdd_2019}.
The PDD is a semi‐physical method that takes the precipitation and the near‐surface air temperature as input and estimates the surface mass balance.
Specifically, the PDD are calculated according to \cite{seguinot_spatial_2013, calov_semi-analytical_2005}
\begin{align}
\mathrm{PDD} \;=\; 
\int_{0}^{A}
\Bigg[
  \frac{\sigma}{\sqrt{2\pi}}
  \exp\!\biggl(-\frac{T_{\mathrm{ac}}(t)^{2}}{2\,\sigma^{2}}\biggr)+\frac{T_{\mathrm{ac}}(t)}{2}\,
  \text{erfc}\!\biggl(-\frac{T_{\mathrm{ac}}(t)}{\sqrt{2}\,\sigma}\biggr)\Bigg]\,\mathrm{d}t,   
\end{align}
with the time interval $A$, the time-dependent seasonal near-surface temperature $T_{\mathrm{ac}}(t)$, the standard deviation of the temperature $\sigma=5\,$K and the error function $\text{erfc}$.
Following the standard conventions, we assume that all precipitation contributes to snow accumulation whenever air temperature $T\leq0^\circ$C, and that the snow‐fraction (hence accumulation) declines linearly to zero as $T$ rises from $0$ to $2^\circ$C. 
Melt is then calculated via positive degree-day factors of $3\,$mm$^\circ$C$^{-1}$d$^{-1}$ for snow and $8\,$mm$^\circ$C$^{-1}$d$^{-1}$ for ice.
For more details we refer to Seguinot et al. (2013) \cite{seguinot_spatial_2013}.
 
\subsection{Training}
We use monthly output from the regional climate model Modèle Atmosphérique Régional (MAR) version 3.12 forced by different Earth system models (Tab.~\ref{tab:models_forcing}) over the time period 1950-2100 \cite{fettweis_estimating_2013, fettweis_reconstructions_2017}.
In total we use 23 different model runs, totaling to 21432 monthly fields. 
We hold out 4 independent runs from MAR forced by the ESM NorESM2 for additional testing.
We regrid the MAR fields to 5\,km resolution, corresponding to the IMSIP6 grid \cite{goelzer_future_2020}.
Additionally, we mask the fields by the current ice sheet mask given by MAR, that is, we set the SMB to 0 outside of the current ice sheet geometry.

We condition our CM on the ice sheet height and the monthly mean insolation as additional input besides the SMB. 
This ensures that generated unconditional samples, i.e., samples without initial conditions, adhere to the topography of the GrIS.
The insolation allows the CM to learn the seasonal melt cycle and generate unconditional samples that adhere to it. 
The input fields are simply concatenated before being fed to the U-Net.

\begin{table}
    \caption{\textbf{Selected MARv3.12 runs for the training forced by CMIP5 and CMIP6 models.} All models used for training and the respective scenario are denoted. The MAR runs forced by NorESM are held out for validation. }
    \centering
    \begin{tabular}{|l|l|}
        \hline
        Forcing (Model) & CMIP Scenarios (runs)\\
        \hline
        ACCESS1.3 & hist, RCP-8.5 \\
        CESM2 & hist, SSP-1.26, SSP-2.45, SSP-5.85\\
        CNRM-CM6 & SSP-5.85\\
        CNRM-ESM2 & hist, SSP-5.85\\
        ERA-5& hist\\
        IPSL-CM6A-LR& hist, SSP-5.85 \\
        MPI-ESM1-2-HR&  hist, SSP-1.26, SSP-2.45, SSP-5.85\\
        UKESM1-0-LL &  hist , SSP-2.45, SSP-5.85 \\ 
        \hline
        NorESM2 & hist, SSP-1.26, SSP-2.45, SSP-5.85\\
     \hline
    \end{tabular}

    \label{tab:models_forcing}
\end{table}

\clearpage

\paragraph{Author Contributions:}{NB conceived and designed the study with input from PH and AR. NB and PH wrote the code. NB analyzed all data. All authors interpreted the results. NB wrote the manuscript with input from AR and PH.} 
\paragraph{Code and Data Availability:}{The original CM code is available at \url{https://github.com/p-hss/consistency-climate-downscaling}. Our modified version for Greenland as well as the training and test data will be made available upon publication.}
\paragraph{Competing Interests:}{The authors declare no competing interests.} 

\paragraph{Acknowledgments:}
{This work was supported by the UiT Aurora Centre Program, UiT - The Arctic University of Norway (2020), and the Research Council of Norway (project number 314570). This is ClimTip contribution \#80; the ClimTip project has received funding from the European Union's Horizon Europe research and innovation programme under grant agreement No. 101137601. Alexander Robinson received funding from the European Union (ERC, FORCLIMA; grant
no. 101044247). The authors gratefully acknowledge the Ministry of Research, Science and Culture (MWFK) of Land Brandenburg for supporting this project by providing resources on the high performance computer system at the Potsdam Institute for Climate Impact Research. The colormaps for the plots are taken from \cite{crameri_misuse_2020}.}

\clearpage

\newpage
\pagenumbering{roman}
\appendix
{\huge Appendix}
\renewcommand\thefigure{\thesection.\arabic{figure}}    
\setcounter{figure}{0}
\section{Unconditional samples}    
\begin{figure}
    \centering
    \includegraphics[width=1\linewidth]{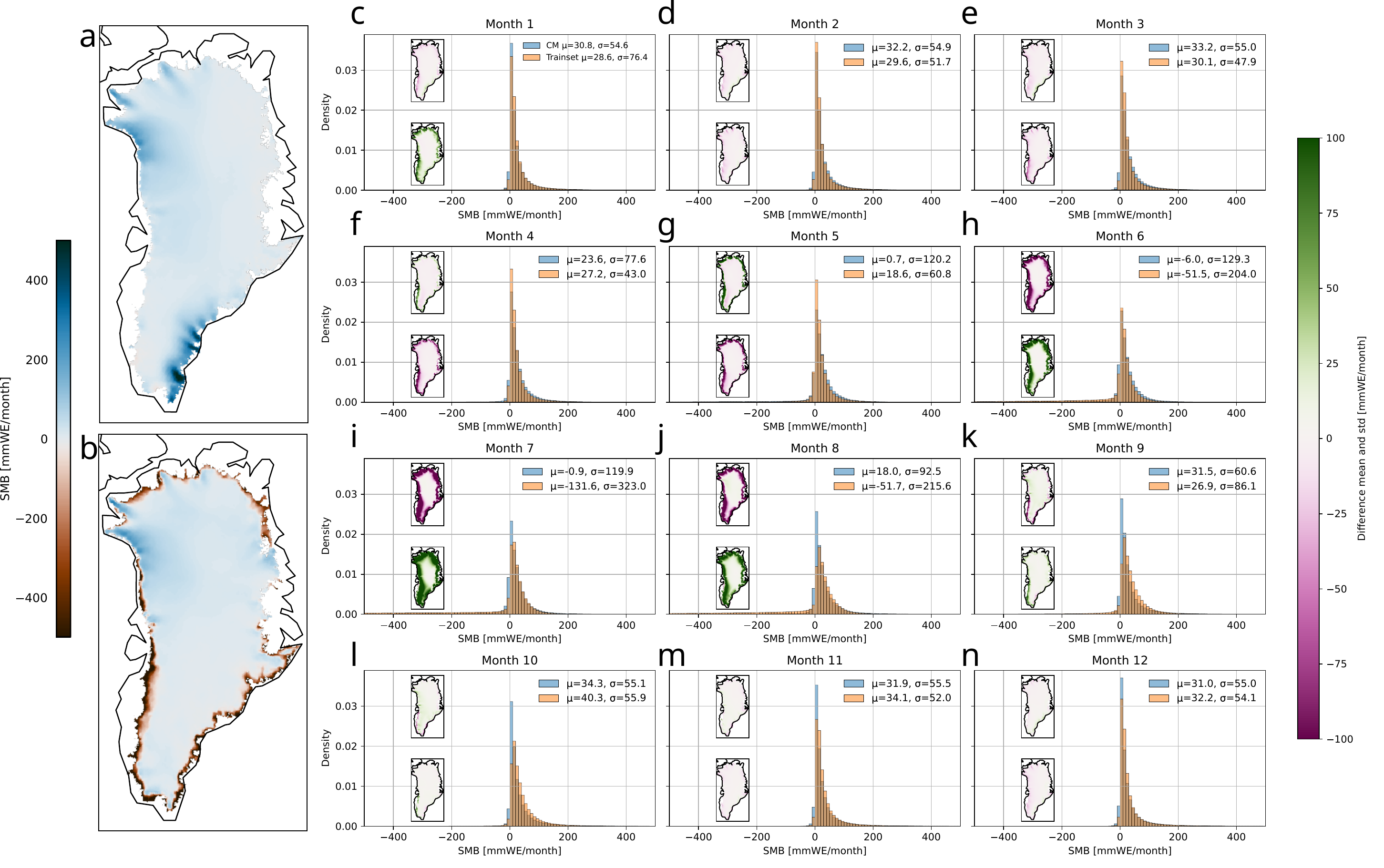}
    \caption{\textbf{Unconditional sampling of SMB from trained CM.}\textbf{(a)} Exemplary unconditional sample of cold month with positive SMB. \textbf{(b)} Exemplary unconditional sample of warm month with negative SMB at the margins. \textbf{(c)} Histogram of generated SMB samples and training set for January. The mean and standard deviation of each set are denoted. The inset shows the spatial difference of the mean (top) and standard deviation (bottom) between generated samples and the training set. Green corresponds to areas where the CM samples have smaller mean/standard deviation, while pink areas denote regions where the CM has larger mean/standard deviation than the training set. \textbf{d-n} Same as \textbf{a} but for all other months of the year. The CM is mostly reproducing the mean and standard deviation well but overestimates the SMB in the summer months with strongly negative SMB.}
    \label{fig:Fig_uncond}
\end{figure}

To show that our models correctly learn the distribution of the underlying training data, we generate 100 unconditional SMB samples for each month and compare them to the training set (Fig.~\ref{fig:Fig_uncond}). 
While, the SMB modeled by MAR is not a ground truth in the classical sense, we treat it as such in lack of observational SMB fields and refer to it as such in the following.
Here, we refer to unconditional samples as samples that are generated from noise, i.e., without any initial guesses of the SMB. 
However, these samples are conditioned on the auxiliary inputs, i.e., the ice sheet height and insolation. 

The CM is able to generate realistic unconditional SMB field samples, both for the winter and summer season (Fig.~\ref{fig:Fig_uncond}a\&b). 
The mean and standard deviation of the generated set closely follow the mean and standard deviation of the training set, except for the summer season (Fig.~\ref{fig:Fig_uncond}h-j). 
In all summer months (JJA), the average SMB is greater than the corresponding training set mean SMB.
Similarly, the standard deviation of the generated ensemble is smaller. 
This is almost exclusively due to the greater SMB at the margins in the generated ensemble.
In other words, the generated ensemble underestimates the melt and variability of the SMB at the ice-sheet margins in the summer months. 
Interestingly, the generated ensemble mean SMB in May is smaller and the variability is greater than that of the training set (Fig.~\ref{fig:Fig_uncond}g).

Generally, we are not interested in generating unconditional samples.
Therefore, the overestimation of the SMB at the margins in the summer months is not a problem for the downscaling task per-se.
We overcome this problem by introducing a hard constraint that conserves the SMB from the low-resolution fields on a regional scale. 
We find, that the hard-constraining generally improves the results. 
Alternatively, oversampling of the summer months during training could improve the unconditional samples for these months.

\clearpage
\section{Examples of downscaled fields}
\setcounter{figure}{0}
\begin{figure}[!htb]
    \centering
    \includegraphics[width=1\linewidth]{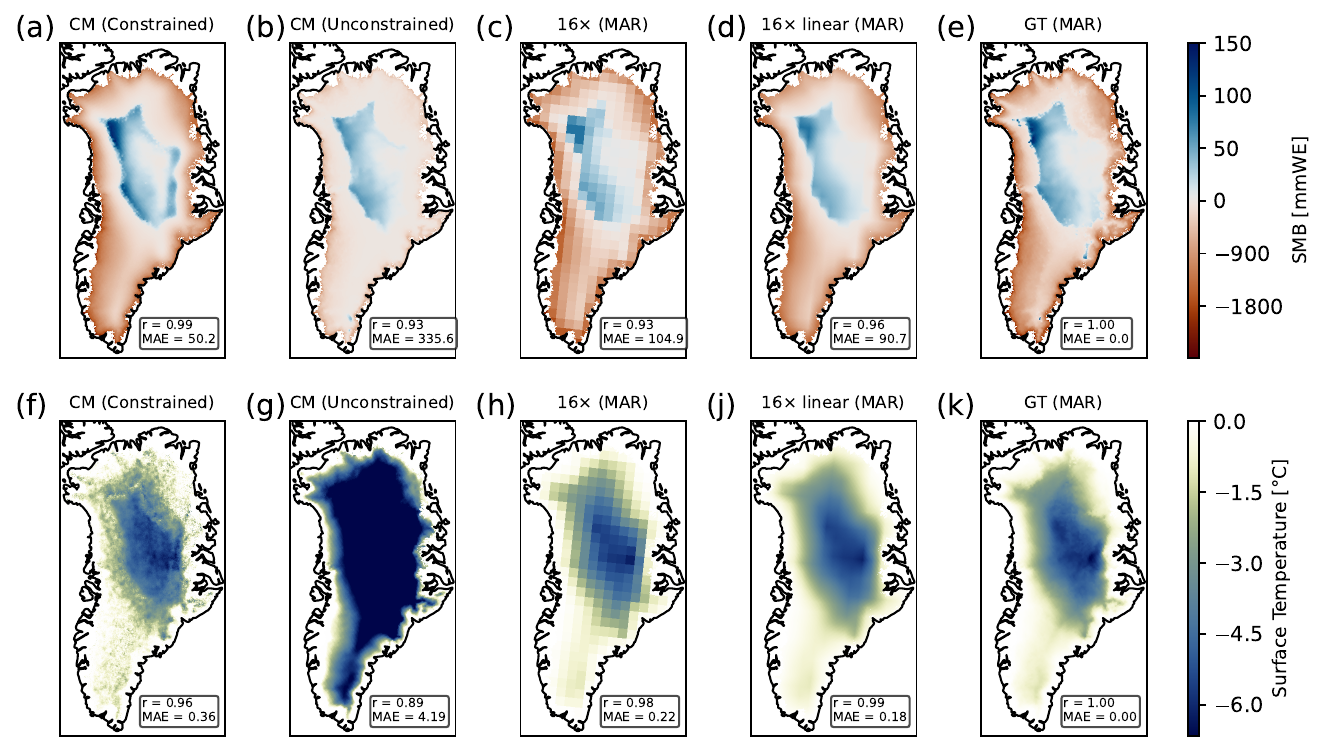}
    \caption{\textbf{Exemplary downscaling of surface mass balance and temperature at surface for random month from test set.} \textbf{(a)} Random month from held-out test set pooled with a factor of 16 (80\,km) resolution. This serves as input to our model. \textbf{(b)} Downscaled SMB field with the CM model using hard constraints to 5\,km resolution. The small scale variability is recovered, especially at the ice-sheet margins. \textbf{(c)} High-resolution ground truth (GT) for this individual month. The GT and the downscaled field are visually indistinguishable. \textbf{(d, e, f)} Same as \textbf{(a, b, c)} but for the temperature at surface. }
    \label{fig:Fig_a1}
\end{figure}
\clearpage

\begin{figure}
    \centering
    \includegraphics[width=0.6\linewidth]{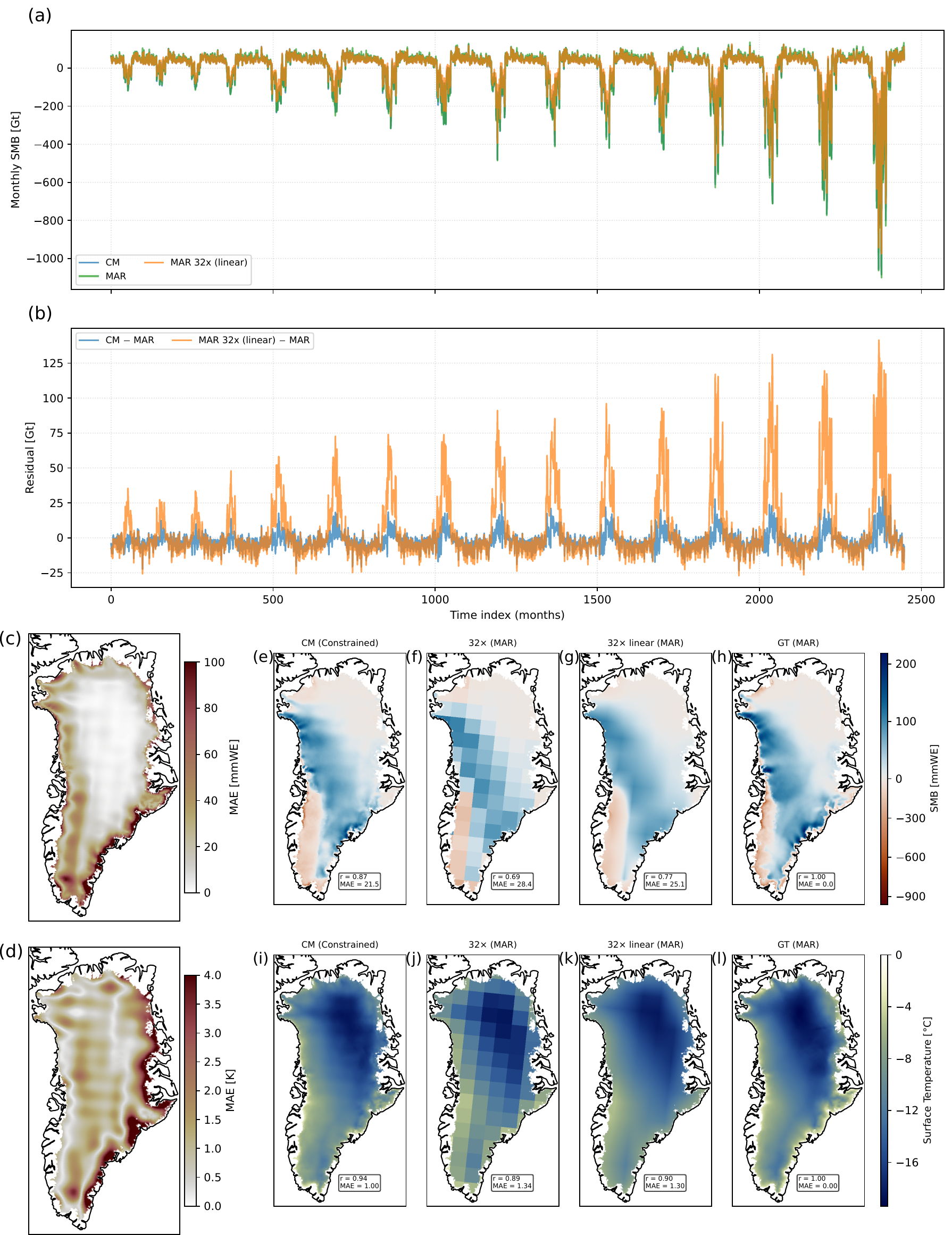}
    \caption{\textbf{Downscaling of coarsened MAR test set by factor 32.}  \textbf{(a)} Monthly integrated SMB over whole test set for MAR, CM and linearly interpolated MAR field from the coarsened MAR field by a factor of 32. \textbf{(b)} Residuals between downscaled field and ground truth MAR and linearly interpolated SMB field and MAR. The linearly interpolated fields overestimates the SMB compared to MAR during the melting seasons. While the downscaled fields also over- and underestimate the SMB they show substantially smaller residuals compared to the linearly interpolated fields.  \textbf{(c)} MAE of downscaled SMB fields over whole test set. The coarse structure of the input field is still partially visible in the MAE field. The average MAE over the whole test set is $24.6\,$mmWE.\textbf{(d)} Same as \textbf{c} but for surface temperature. The MAE over the whole test set is $1.35\,$K. \textbf{(e)}  Downscaled SMB field of test set with hard constraints during inference for warm month with negative SMB at the margins (blue) with a 5\,km resolution. The mean absolute error (in mmWE) and Pearson correlation with the GT field are denoted. \textbf{(b)} Coarsened MAR field by a factor of 32 (160\,km resolution), which is the input for the CM. The small scale variability is lost. \textbf{(d)} Same as \textbf{b} but linearly interpolated. 
    \textbf{(e)} Ground truth MAR field with 5\,km resolution. 
    \textbf{(f-j)} Same as \textbf{e-h} but for the surface temperature.}
    \label{fig:Fig_32x}
\end{figure}
\clearpage

\section{NorESM2 downscaling}
\setcounter{figure}{0}

\begin{figure}[!htb]
    \centering
    \includegraphics[width=1\linewidth]{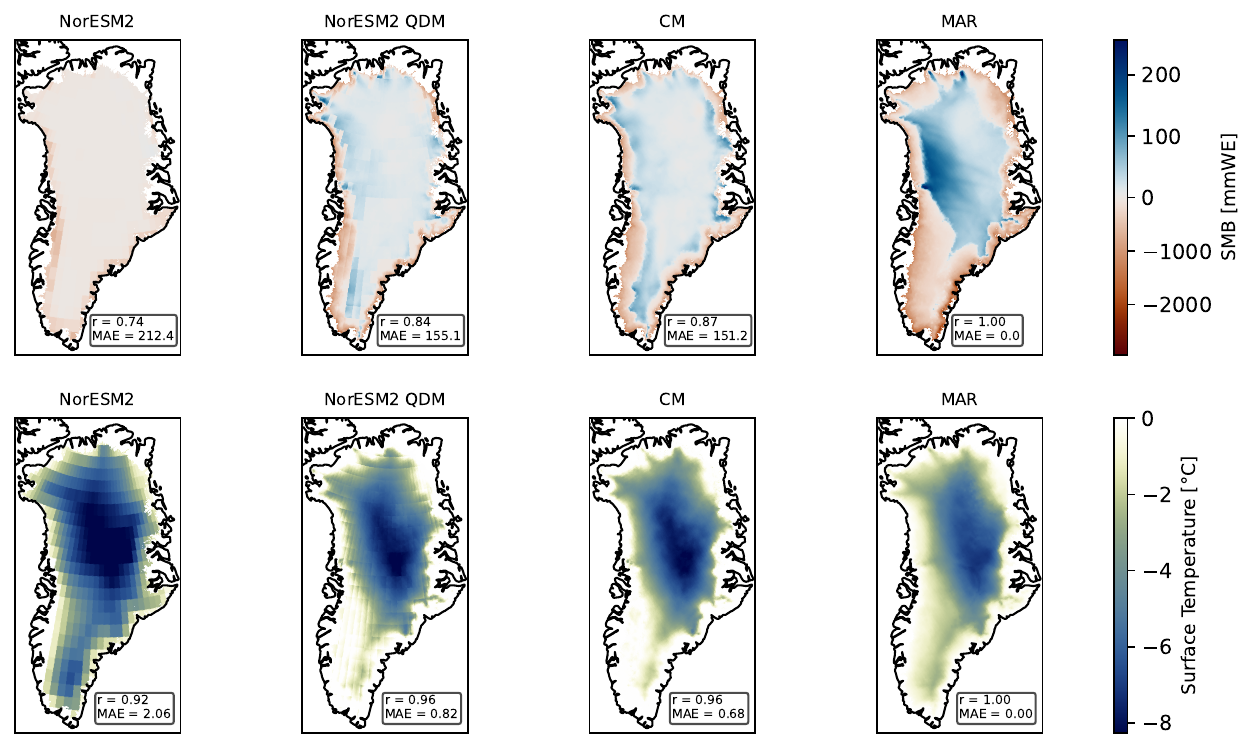}
    \caption{\textbf{Exemplary downscaling of surface mass balance and temperature at surface for end of century month for NorESM2 SMB.} \textbf{(a)} NorESM2 SMB as sum of precipitation, melt and runoff. The SMB does not show any strong negative SMB at the margins. The MAE and correlation compared to the MAR SMB for the same month are denoted. \textbf{(b)} QDM-corrected NorESM2 SMB. The margins show a more negative SMB compared to non-bias corrected SMB. However, the SMB is still overestimated compared to the MAR SMB. \textbf{{c}} Downscaled SMB field with the CM model using hard constraints to 5\,km resolution. \textbf{{d}} Unpaired high-resolution ground truth (GT) for this individual month as simulated by MAR. \textbf{(e, f, g, h)} Same as \textbf{(a, b, c, d)} but for the temperature at surface. The bias-corrected and downscaled $T_s$ show stronger similarity with MAR than the SMB fields.}
    \label{fig:Fig_noresm_qdmwarm}
\end{figure}
\clearpage

\begin{figure}
    \centering
    \includegraphics[width=1\linewidth]{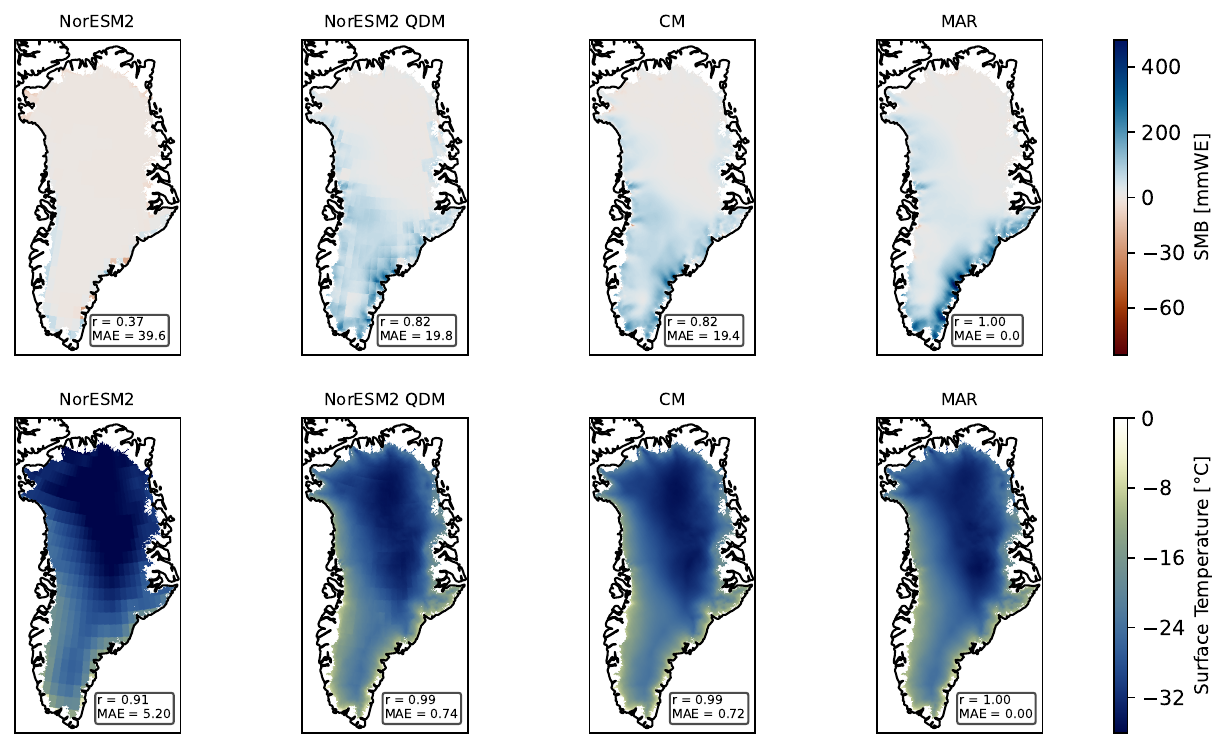}
    \caption{\textbf{Exemplary downscaling of surface mass balance and temperature at surface for random month from test set.} \textbf{(a)} Random month from held-out test set pooled with a factor of 16 (80\,km) resolution. This serves as input to our model. \textbf{(b)} Downscaled SMB field with the CM model using hard constraints to 5\,km resolution. The small scale variability is recovered, especially at the ice-sheet margins. \textbf{(c)} High-resolution ground truth (GT) for this individual month. The GT and the downscaled field are visually indistinguishable. \textbf{(d, e, f)} Same as \textbf{(a, b, c)} but for the temperature at surface. }
    \label{fig:Fig_noresm_qdmcold}
\end{figure}
\clearpage

\begin{figure}
    \centering
    \includegraphics[width=1\linewidth]{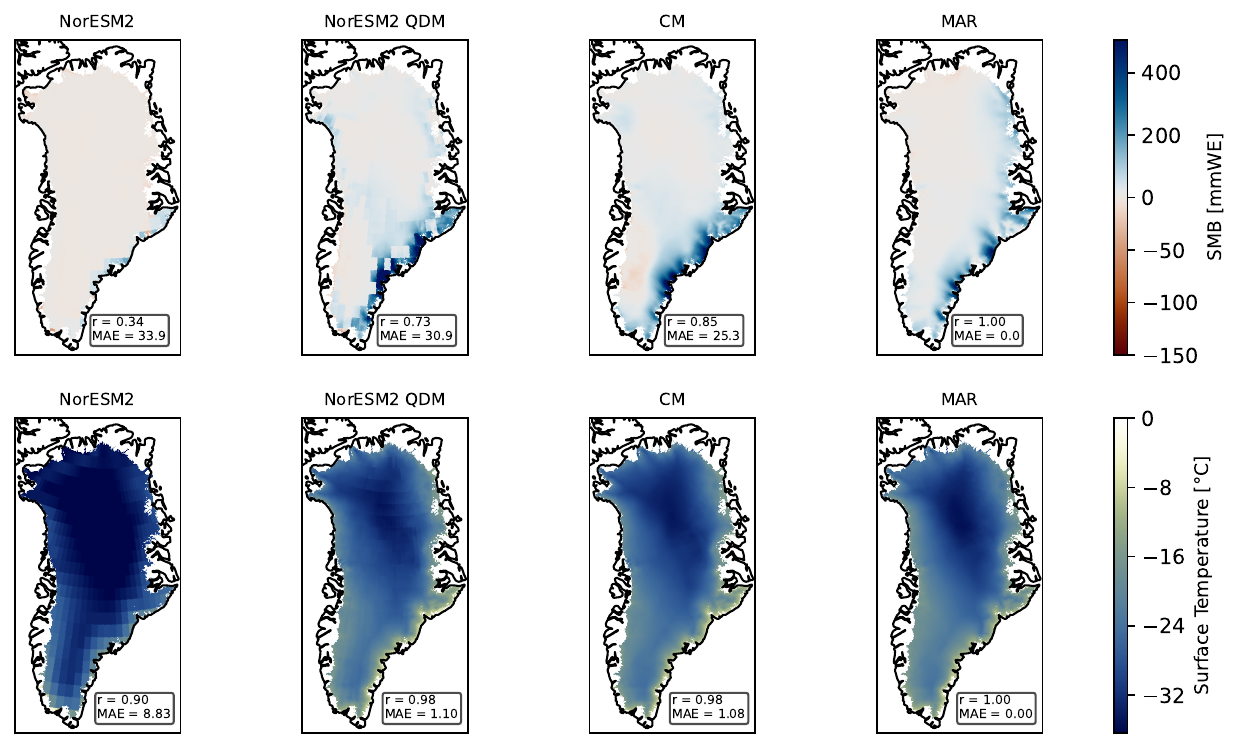}
    \caption{\textbf{Exemplary downscaling of surface mass balance and temperature at surface for random month from test set.} \textbf{(a)} Random month from held-out test set pooled with a factor of 16 (80\,km) resolution. This serves as input to our model. \textbf{(b)} Downscaled SMB field with the CM model using hard constraints to 5\,km resolution. The small scale variability is recovered, especially at the ice-sheet margins. \textbf{(c)} High-resolution ground truth (GT) for this individual month. The GT and the downscaled field are visually indistinguishable. \textbf{(d, e, f)} Same as \textbf{(a, b, c)} but for the temperature at surface. }
    \label{fig:Fig_noresm_qdm_wrong}
\end{figure}
\clearpage

\begin{figure}
    \centering
    \includegraphics[width=1\linewidth]{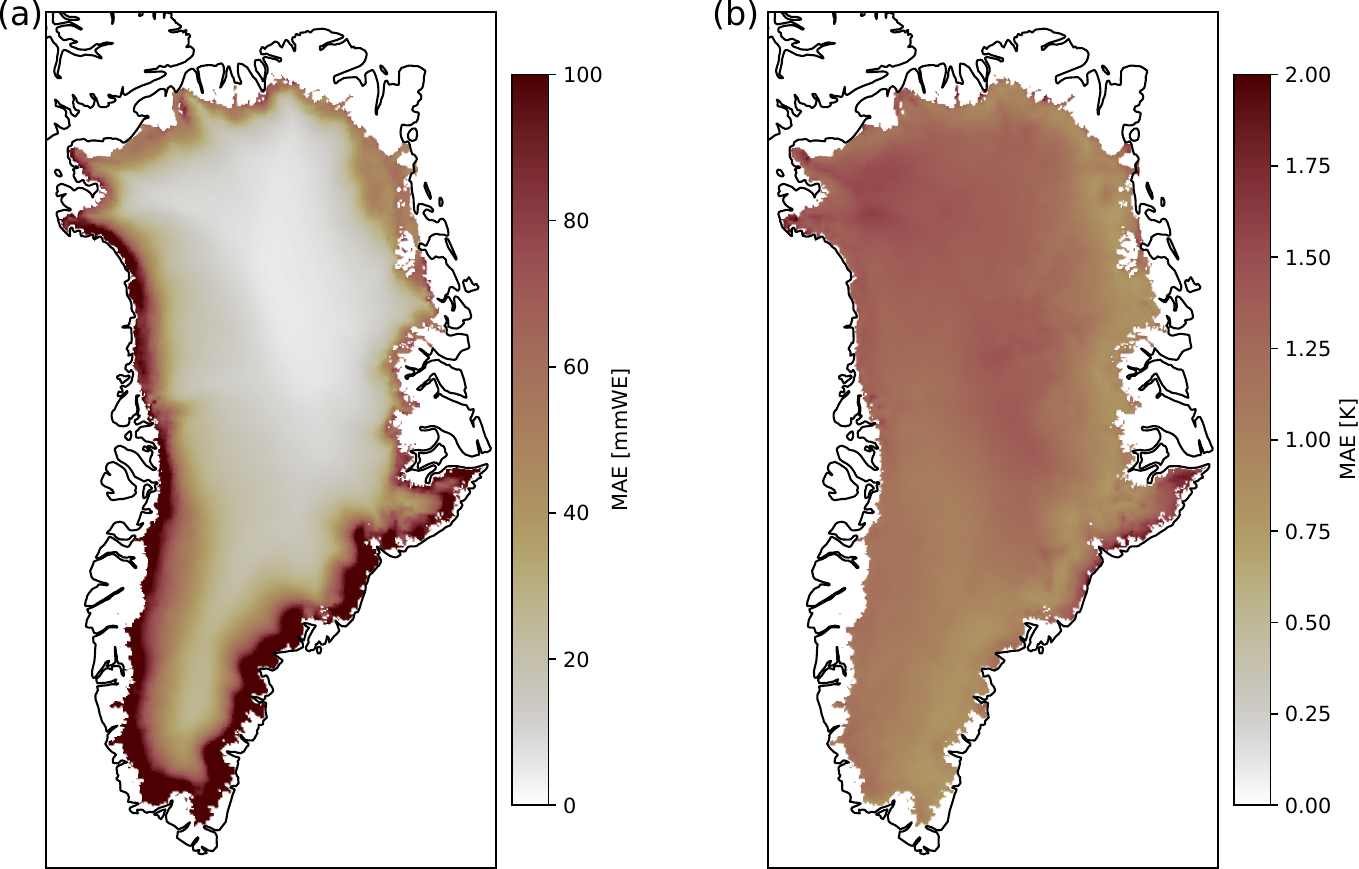}
    \caption{\textbf{Mean absolute error over whole time period (2015-2100) for QDM-corrected NorESM2 SMB for SSP-5.85 scenario.} \textbf{(a)} Temporal MAE for each grid cell of downscaled SMB from bias-corrected SMB field derived from NorESM2 SSP-5.85 run. The margins show the largest error. \textbf{(b)} Same as \textbf{a} but for surface temperature.  }
    \label{fig:Fig_noresm_qdm_mae}
\end{figure}
\clearpage

\begin{figure}
    \centering
    \includegraphics[width=0.7\linewidth]{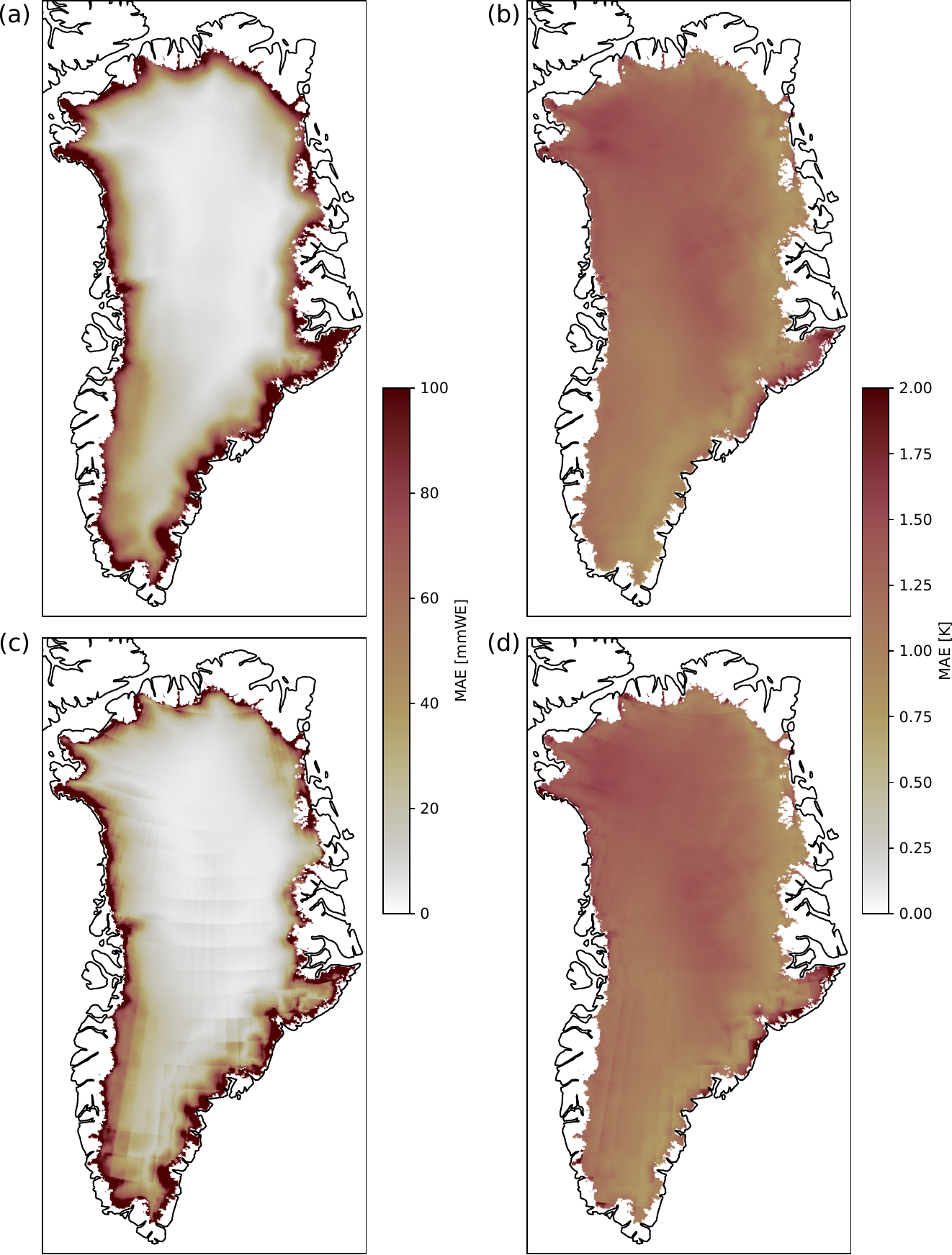}
    \caption{\textbf{Mean absolute error over whole time period (2015-2100) for PDD NorESM2 SMB for SSP-5.85 scenario.} \textbf{(a)} Temporal MAE for each grid cell of downscaled SMB from bias-corrected PDD derived SMB field forced by NorESM2 SSP-5.85 run. The margins show the largest error. \textbf{(b)} Same as \textbf{a} but for surface temperature. \textbf{(c,d)} Same as \textbf{a,b} but directly for bias corrected SMB and $T_s$ fields without downscaling.}
    \label{fig:Fig_noresm_pdd_mae}
\end{figure}

\clearpage

 \section{Training and validation loss}
 \setcounter{figure}{0}
\begin{figure}[!htb]
    \centering
    \includegraphics[width=1\linewidth]{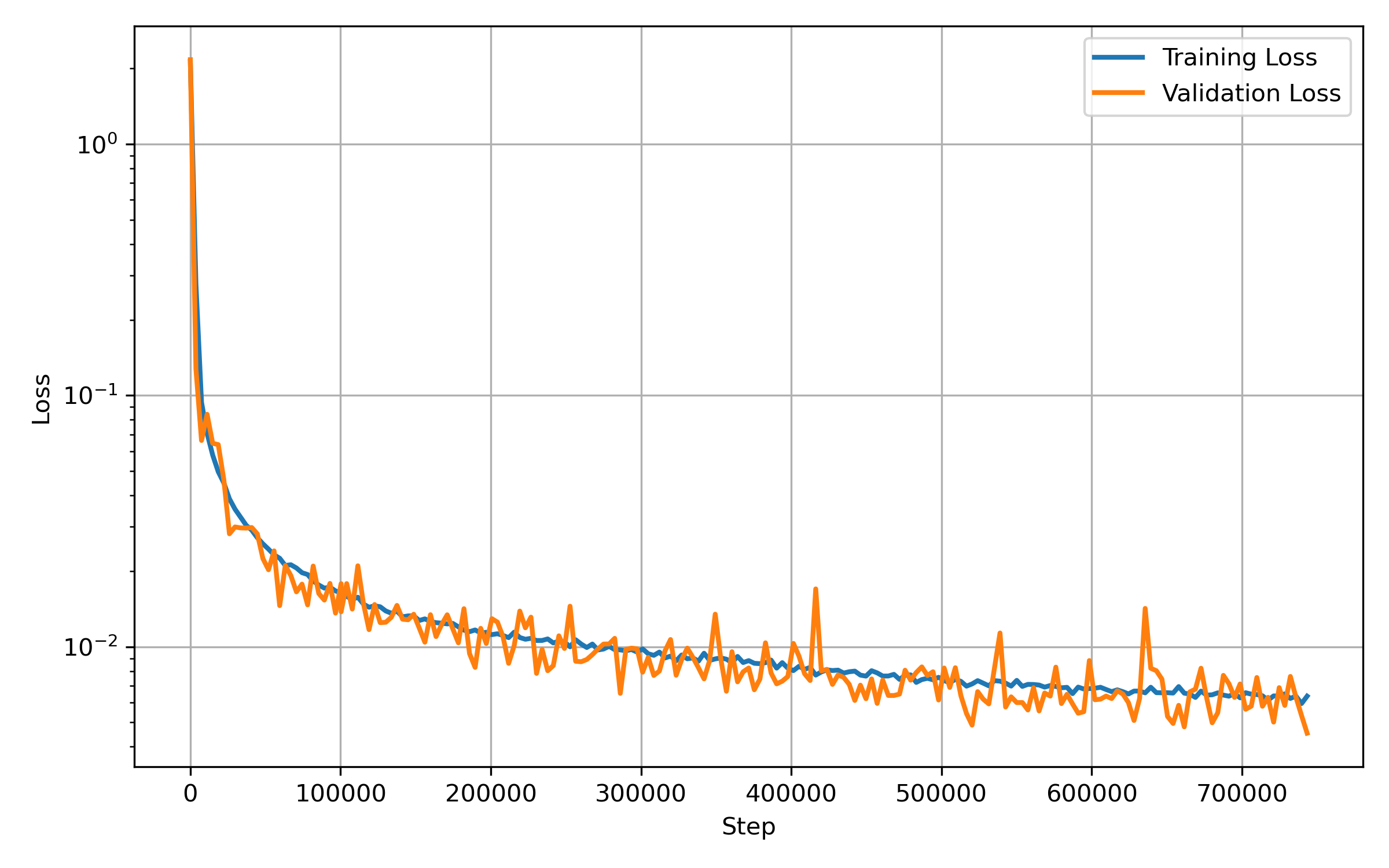}
    \caption{\textbf{Validation and training loss of consistency model.}  }
    \label{fig:Fig_loss}
\end{figure}

\newpage

\clearpage

\printbibliography

@article{bourgault_xclim_2023,
	title = {xclim: xarray-based climate data analytics},
	volume = {8},
	issn = {2475-9066},
	shorttitle = {xclim},
	url = {https://joss.theoj.org/papers/10.21105/joss.05415},
	doi = {10.21105/joss.05415},
	language = {en},
	number = {85},
	urldate = {2025-03-04},
	journal = {Journal of Open Source Software},
	author = {Bourgault, Pascal and Huard, David and Smith, Trevor James and Logan, Travis and Aoun, Abel and Lavoie, Juliette and Dupuis, Eric and Rondeau-Genesse, Gabriel and Alegre, Raquel and Barnes, Clair and Laperrière, Alexis Beaupré and Biner, Sébastien and Caron, David and Ehbrecht, Carsten and Fyke, Jeremy and Keel, Tom and Labonté, Marie-Pier and Lierhammer, Ludwig and Low, Jwen-Fai and Quinn, Jamie and Roy, Philippe and Squire, Dougie and Stephens, Ag and Tanguy, Maliko and Whelan, Christopher},
	month = may,
	year = {2023},
	pages = {5415},
}

@article{mottram_surface_2017,
	title = {Surface mass balance of the {Greenland} ice sheet in the regional climate model {HIRHAM5}: {Present} state and future prospects},
	volume = {75},
	shorttitle = {Surface mass balance of the {Greenland} ice sheet in the regional climate model {HIRHAM5}},
	url = {https://eprints.lib.hokudai.ac.jp/dspace/handle/2115/65106},
	urldate = {2025-07-22},
	journal = {Low Temperature Science},
	author = {Mottram, Ruth and Boberg, Fredrik and Langen, Peter and Yang, Shuting and Rodehacke, Christian and Christensen, Jens Hesselbjerg and Madsen, Marianne Sloth},
	month = mar,
	year = {2017},
	pages = {105--115},
}

@misc{aich_conditional_2024,
	title = {Conditional diffusion models for downscaling \& bias correction of {Earth} system model precipitation},
	url = {http://arxiv.org/abs/2404.14416},
	doi = {10.48550/arXiv.2404.14416},
	urldate = {2025-07-23},
	publisher = {arXiv},
	author = {Aich, Michael and Hess, Philipp and Pan, Baoxiang and Bathiany, Sebastian and Huang, Yu and Boers, Niklas},
	month = apr,
	year = {2024},
	note = {arXiv:2404.14416 [physics]},
}

@misc{ho_denoising_2020,
	title = {Denoising {Diffusion} {Probabilistic} {Models}},
	url = {http://arxiv.org/abs/2006.11239},
	doi = {10.48550/arXiv.2006.11239},
	urldate = {2025-07-22},
	publisher = {arXiv},
	author = {Ho, Jonathan and Jain, Ajay and Abbeel, Pieter},
	month = dec,
	year = {2020},
	note = {arXiv:2006.11239 [cs]},
}

@misc{song_score-based_2021,
	title = {Score-{Based} {Generative} {Modeling} through {Stochastic} {Differential} {Equations}},
	url = {http://arxiv.org/abs/2011.13456},
	doi = {10.48550/arXiv.2011.13456},
	urldate = {2025-07-22},
	publisher = {arXiv},
	author = {Song, Yang and Sohl-Dickstein, Jascha and Kingma, Diederik P. and Kumar, Abhishek and Ermon, Stefano and Poole, Ben},
	month = feb,
	year = {2021},
	note = {arXiv:2011.13456 [cs]},
}

@article{goelzer_future_2020,
	title = {The future sea-level contribution of the {Greenland} ice sheet: a multi-model ensemble study of {ISMIP6}},
	volume = {14},
	issn = {1994-0416},
	shorttitle = {The future sea-level contribution of the {Greenland} ice sheet},
	url = {https://tc.copernicus.org/articles/14/3071/2020/},
	doi = {10.5194/tc-14-3071-2020},
	language = {English},
	number = {9},
	urldate = {2025-07-22},
	journal = {The Cryosphere},
	author = {Goelzer, Heiko and Nowicki, Sophie and Payne, Anthony and Larour, Eric and Seroussi, Helene and Lipscomb, William H. and Gregory, Jonathan and Abe-Ouchi, Ayako and Shepherd, Andrew and Simon, Erika and Agosta, Cécile and Alexander, Patrick and Aschwanden, Andy and Barthel, Alice and Calov, Reinhard and Chambers, Christopher and Choi, Youngmin and Cuzzone, Joshua and Dumas, Christophe and Edwards, Tamsin and Felikson, Denis and Fettweis, Xavier and Golledge, Nicholas R. and Greve, Ralf and Humbert, Angelika and Huybrechts, Philippe and Le clec'h, Sebastien and Lee, Victoria and Leguy, Gunter and Little, Chris and Lowry, Daniel P. and Morlighem, Mathieu and Nias, Isabel and Quiquet, Aurelien and Rückamp, Martin and Schlegel, Nicole-Jeanne and Slater, Donald A. and Smith, Robin S. and Straneo, Fiamma and Tarasov, Lev and van de Wal, Roderik and van den Broeke, Michiel},
	month = sep,
	year = {2020},
	note = {Publisher: Copernicus GmbH},
	pages = {3071--3096},
}

@article{noel_daily_2016,
	title = {A daily, 1 km resolution data set of downscaled {Greenland} ice sheet surface mass balance (1958–2015)},
	volume = {10},
	issn = {1994-0416},
	url = {https://tc.copernicus.org/articles/10/2361/2016/},
	doi = {10.5194/tc-10-2361-2016},
	language = {English},
	number = {5},
	urldate = {2025-07-22},
	journal = {The Cryosphere},
	author = {Noël, Brice and van de Berg, Willem Jan and Machguth, Horst and Lhermitte, Stef and Howat, Ian and Fettweis, Xavier and van den Broeke, Michiel R.},
	month = oct,
	year = {2016},
	note = {Publisher: Copernicus GmbH},
	pages = {2361--2377},
}

@article{van_den_broeke_recent_2016,
	title = {On the recent contribution of the {Greenland} ice sheet to sea level change},
	volume = {10},
	issn = {1994-0416},
	url = {https://tc.copernicus.org/articles/10/1933/2016/},
	doi = {10.5194/tc-10-1933-2016},
	language = {English},
	number = {5},
	urldate = {2025-07-22},
	journal = {The Cryosphere},
	author = {van den Broeke, Michiel R. and Enderlin, Ellyn M. and Howat, Ian M. and Kuipers Munneke, Peter and Noël, Brice P. Y. and van de Berg, Willem Jan and van Meijgaard, Erik and Wouters, Bert},
	month = sep,
	year = {2016},
	note = {Publisher: Copernicus GmbH},
	pages = {1933--1946},
}

@article{lucas-picher_very_2012,
	title = {Very high resolution regional climate model simulations over {Greenland}: {Identifying} added value},
	volume = {117},
	copyright = {Copyright 2012 by the American Geophysical Union},
	issn = {2156-2202},
	shorttitle = {Very high resolution regional climate model simulations over {Greenland}},
	url = {https://onlinelibrary.wiley.com/doi/abs/10.1029/2011JD016267},
	doi = {10.1029/2011JD016267},
	language = {en},
	number = {D2},
	urldate = {2025-07-22},
	journal = {Journal of Geophysical Research: Atmospheres},
	author = {Lucas-Picher, Philippe and Wulff-Nielsen, Maria and Christensen, Jens H. and Aðalgeirsdóttir, Guðfinna and Mottram, Ruth and Simonsen, Sebastian B.},
	year = {2012},
	note = {\_eprint: https://agupubs.onlinelibrary.wiley.com/doi/pdf/10.1029/2011JD016267},
}

@article{watt-meyer_ace2_2025,
	title = {{ACE2}: accurately learning subseasonal to decadal atmospheric variability and forced responses},
	volume = {8},
	copyright = {2025 The Author(s)},
	issn = {2397-3722},
	shorttitle = {{ACE2}},
	url = {https://www.nature.com/articles/s41612-025-01090-0},
	doi = {10.1038/s41612-025-01090-0},
	language = {en},
	number = {1},
	urldate = {2025-07-22},
	journal = {npj Climate and Atmospheric Science},
	author = {Watt-Meyer, Oliver and Henn, Brian and McGibbon, Jeremy and Clark, Spencer K. and Kwa, Anna and Perkins, W. Andre and Wu, Elynn and Harris, Lucas and Bretherton, Christopher S.},
	month = may,
	year = {2025},
	note = {Publisher: Nature Publishing Group},
	pages = {205},
}

@article{price_probabilistic_2025,
	title = {Probabilistic weather forecasting with machine learning},
	volume = {637},
	copyright = {2024 The Author(s)},
	issn = {1476-4687},
	url = {https://www.nature.com/articles/s41586-024-08252-9},
	doi = {10.1038/s41586-024-08252-9},
	language = {en},
	number = {8044},
	urldate = {2025-07-22},
	journal = {Nature},
	author = {Price, Ilan and Sanchez-Gonzalez, Alvaro and Alet, Ferran and Andersson, Tom R. and El-Kadi, Andrew and Masters, Dominic and Ewalds, Timo and Stott, Jacklynn and Mohamed, Shakir and Battaglia, Peter and Lam, Remi and Willson, Matthew},
	month = jan,
	year = {2025},
	note = {Publisher: Nature Publishing Group},
	pages = {84--90},
}

@article{crameri_misuse_2020,
	title = {The misuse of colour in science communication},
	volume = {11},
	copyright = {2020 The Author(s)},
	issn = {2041-1723},
	url = {https://www.nature.com/articles/s41467-020-19160-7},
	doi = {10.1038/s41467-020-19160-7},
	language = {en},
	number = {1},
	urldate = {2022-12-13},
	journal = {Nature Communications},
	author = {Crameri, Fabio and Shephard, Grace E. and Heron, Philip J.},
	month = oct,
	year = {2020},
	note = {Number: 1
Publisher: Nature Publishing Group},
	pages = {5444},
}

@article{delhasse_coupling_2024,
	title = {Coupling {MAR} ({Modèle} {Atmosphérique} {Régional}) with {PISM} ({Parallel} {Ice} {Sheet} {Model}) mitigates the positive melt–elevation feedback},
	volume = {18},
	issn = {1994-0416},
	url = {https://tc.copernicus.org/articles/18/633/2024/},
	doi = {10.5194/tc-18-633-2024},
	language = {English},
	number = {2},
	urldate = {2025-06-19},
	journal = {The Cryosphere},
	author = {Delhasse, Alison and Beckmann, Johanna and Kittel, Christoph and Fettweis, Xavier},
	month = feb,
	year = {2024},
	note = {Publisher: Copernicus GmbH},
	pages = {633--651},
}

@article{sellevold_surface_2019,
	title = {Surface mass balance downscaling through elevation classes in an {Earth} system model: application to the {Greenland} ice sheet},
	volume = {13},
	issn = {1994-0416},
	shorttitle = {Surface mass balance downscaling through elevation classes in an {Earth} system model},
	url = {https://tc.copernicus.org/articles/13/3193/2019/},
	doi = {10.5194/tc-13-3193-2019},
	language = {English},
	number = {12},
	urldate = {2025-06-19},
	journal = {The Cryosphere},
	author = {Sellevold, Raymond and van Kampenhout, Leonardus and Lenaerts, Jan T. M. and Noël, Brice and Lipscomb, William H. and Vizcaino, Miren},
	month = dec,
	year = {2019},
	note = {Publisher: Copernicus GmbH},
	pages = {3193--3208},
}

@article{bi_accurate_2023,
	title = {Accurate medium-range global weather forecasting with {3D} neural networks},
	volume = {619},
	copyright = {2023 The Author(s)},
	issn = {1476-4687},
	url = {https://www.nature.com/articles/s41586-023-06185-3},
	doi = {10.1038/s41586-023-06185-3},
	language = {en},
	number = {7970},
	urldate = {2025-06-19},
	journal = {Nature},
	author = {Bi, Kaifeng and Xie, Lingxi and Zhang, Hengheng and Chen, Xin and Gu, Xiaotao and Tian, Qi},
	month = jul,
	year = {2023},
	note = {Publisher: Nature Publishing Group},
	pages = {533--538},
}

@article{payne_future_2021,
	title = {Future {Sea} {Level} {Change} {Under} {Coupled} {Model} {Intercomparison} {Project} {Phase} 5 and {Phase} 6 {Scenarios} {From} the {Greenland} and {Antarctic} {Ice} {Sheets}},
	volume = {48},
	copyright = {© 2021. The Authors.},
	issn = {1944-8007},
	url = {https://onlinelibrary.wiley.com/doi/abs/10.1029/2020GL091741},
	doi = {10.1029/2020GL091741},
	language = {en},
	number = {16},
	urldate = {2024-04-04},
	journal = {Geophysical Research Letters},
	author = {Payne, Antony J. and Nowicki, Sophie and Abe-Ouchi, Ayako and Agosta, Cécile and Alexander, Patrick and Albrecht, Torsten and Asay-Davis, Xylar and Aschwanden, Andy and Barthel, Alice and Bracegirdle, Thomas J. and Calov, Reinhard and Chambers, Christopher and Choi, Youngmin and Cullather, Richard and Cuzzone, Joshua and Dumas, Christophe and Edwards, Tamsin L. and Felikson, Denis and Fettweis, Xavier and Galton-Fenzi, Benjamin K. and Goelzer, Heiko and Gladstone, Rupert and Golledge, Nicholas R. and Gregory, Jonathan M. and Greve, Ralf and Hattermann, Tore and Hoffman, Matthew J. and Humbert, Angelika and Huybrechts, Philippe and Jourdain, Nicolas C. and Kleiner, Thomas and Munneke, Peter Kuipers and Larour, Eric and Le clec'h, Sebastien and Lee, Victoria and Leguy, Gunter and Lipscomb, William H. and Little, Christopher M. and Lowry, Daniel P. and Morlighem, Mathieu and Nias, Isabel and Pattyn, Frank and Pelle, Tyler and Price, Stephen F. and Quiquet, Aurélien and Reese, Ronja and Rückamp, Martin and Schlegel, Nicole-Jeanne and Seroussi, Hélène and Shepherd, Andrew and Simon, Erika and Slater, Donald and Smith, Robin S. and Straneo, Fiammetta and Sun, Sainan and Tarasov, Lev and Trusel, Luke D. and Van Breedam, Jonas and van de Wal, Roderik and van den Broeke, Michiel and Winkelmann, Ricarda and Zhao, Chen and Zhang, Tong and Zwinger, Thomas},
	year = {2021},
	note = {\_eprint: https://onlinelibrary.wiley.com/doi/pdf/10.1029/2020GL091741},
	pages = {e2020GL091741},
}

@article{trusel_nonlinear_2018,
	title = {Nonlinear rise in {Greenland} runoff in response to post-industrial {Arctic} warming},
	volume = {564},
	copyright = {2018 Springer Nature Limited},
	issn = {1476-4687},
	url = {https://www.nature.com/articles/s41586-018-0752-4},
	doi = {10.1038/s41586-018-0752-4},
	language = {en},
	number = {7734},
	urldate = {2025-06-18},
	journal = {Nature},
	author = {Trusel, Luke D. and Das, Sarah B. and Osman, Matthew B. and Evans, Matthew J. and Smith, Ben E. and Fettweis, Xavier and McConnell, Joseph R. and Noël, Brice P. Y. and van den Broeke, Michiel R.},
	month = dec,
	year = {2018},
	note = {Publisher: Nature Publishing Group},
	pages = {104--108},
}

@article{morlighem_bedmachine_2017,
	title = {{BedMachine} v3: {Complete} {Bed} {Topography} and {Ocean} {Bathymetry} {Mapping} of {Greenland} {From} {Multibeam} {Echo} {Sounding} {Combined} {With} {Mass} {Conservation}},
	volume = {44},
	copyright = {©2017. The Authors.},
	issn = {1944-8007},
	shorttitle = {{BedMachine} v3},
	url = {https://onlinelibrary.wiley.com/doi/abs/10.1002/2017GL074954},
	doi = {10.1002/2017GL074954},
	language = {en},
	number = {21},
	urldate = {2025-06-18},
	journal = {Geophysical Research Letters},
	author = {Morlighem, M. and Williams, C. N. and Rignot, E. and An, L. and Arndt, J. E. and Bamber, J. L. and Catania, G. and Chauché, N. and Dowdeswell, J. A. and Dorschel, B. and Fenty, I. and Hogan, K. and Howat, I. and Hubbard, A. and Jakobsson, M. and Jordan, T. M. and Kjeldsen, K. K. and Millan, R. and Mayer, L. and Mouginot, J. and Noël, B. P. Y. and O'Cofaigh, C. and Palmer, S. and Rysgaard, S. and Seroussi, H. and Siegert, M. J. and Slabon, P. and Straneo, F. and van den Broeke, M. R. and Weinrebe, W. and Wood, M. and Zinglersen, K. B.},
	year = {2017},
	note = {\_eprint: https://agupubs.onlinelibrary.wiley.com/doi/pdf/10.1002/2017GL074954},
	pages = {11,051--11,061},
}

@article{van_dalum_first_2024,
	title = {First results of the polar regional climate model {RACMO2}.4},
	volume = {18},
	issn = {1994-0416},
	url = {https://tc.copernicus.org/articles/18/4065/2024/},
	doi = {10.5194/tc-18-4065-2024},
	language = {English},
	number = {9},
	urldate = {2025-06-17},
	journal = {The Cryosphere},
	author = {van Dalum, Christiaan T. and van de Berg, Willem Jan and Gadde, Srinidhi N. and van Tiggelen, Maurice and van der Drift, Tijmen and van Meijgaard, Erik and van Ulft, Lambertus H. and van den Broeke, Michiel R.},
	month = sep,
	year = {2024},
	note = {Publisher: Copernicus GmbH},
	pages = {4065--4088},
}

@article{tedesco_computationally_2023,
	title = {A computationally efficient statistically downscaled 100\&thinsp;m resolution {Greenland} product from the regional climate model {MAR}},
	volume = {17},
	issn = {1994-0416},
	url = {https://tc.copernicus.org/articles/17/5061/2023/},
	doi = {10.5194/tc-17-5061-2023},
	language = {English},
	number = {12},
	urldate = {2025-06-17},
	journal = {The Cryosphere},
	author = {Tedesco, Marco and Colosio, Paolo and Fettweis, Xavier and Cervone, Guido},
	month = nov,
	year = {2023},
	note = {Publisher: Copernicus GmbH},
	pages = {5061--5074},
}

@article{calov_semi-analytical_2005,
	title = {A semi-analytical solution for the positive degree-day model with stochastic temperature variations},
	volume = {51},
	issn = {0022-1430, 1727-5652},
	url = {https://www.cambridge.org/core/journals/journal-of-glaciology/article/semianalytical-solution-for-the-positive-degreeday-model-with-stochastic-temperature-variations/6FD568E1D92AA7B21C25E34B012790E1},
	doi = {10.3189/172756505781829601},
	language = {en},
	number = {172},
	urldate = {2025-06-17},
	journal = {Journal of Glaciology},
	author = {Calov, Reinhard and Greve, Ralf},
	month = jan,
	year = {2005},
	pages = {173--175},
}

@article{seguinot_spatial_2013,
	title = {Spatial and seasonal effects of temperature variability in a positive degree-day glacier surface mass-balance model},
	volume = {59},
	issn = {0022-1430, 1727-5652},
	url = {https://www.cambridge.org/core/journals/journal-of-glaciology/article/spatial-and-seasonal-effects-of-temperature-variability-in-a-positive-degreeday-glacier-surface-massbalance-model/79103AC980965B2499CFFD4C6D10CC3A},
	doi = {10.3189/2013JoG13J081},
	language = {en},
	number = {218},
	urldate = {2025-06-17},
	journal = {Journal of Glaciology},
	author = {Seguinot, Julien},
	month = jan,
	year = {2013},
	pages = {1202--1204},
}

@misc{seguinot_pypdd_2019,
	title = {{PyPDD}: a positive degree day model for glacier surface mass balance},
	shorttitle = {{PyPDD}},
	url = {https://zenodo.org/records/3467639},
	urldate = {2025-06-17},
	publisher = {Zenodo},
	author = {Seguinot, Julien},
	month = oct,
	year = {2019},
	doi = {10.5281/zenodo.3467639},
}

@article{bischoff_unpaired_2024,
	title = {Unpaired {Downscaling} of {Fluid} {Flows} with {Diffusion} {Bridges}},
	volume = {3},
	issn = {2769-7525},
	url = {https://journals.ametsoc.org/view/journals/aies/3/2/AIES-D-23-0039.1.xml},
	doi = {10.1175/AIES-D-23-0039.1},
	language = {EN},
	number = {2},
	urldate = {2025-06-16},
	journal = {Artificial Intelligence for the Earth Systems},
	author = {Bischoff, Tobias and Deck, Katherine},
	month = may,
	year = {2024},
	note = {Publisher: American Meteorological Society
Section: Artificial Intelligence for the Earth Systems},
}

@misc{arjovsky_towards_2017,
	title = {Towards {Principled} {Methods} for {Training} {Generative} {Adversarial} {Networks}},
	url = {http://arxiv.org/abs/1701.04862},
	doi = {10.48550/arXiv.1701.04862},
	urldate = {2025-06-16},
	publisher = {arXiv},
	author = {Arjovsky, Martin and Bottou, Léon},
	month = jan,
	year = {2017},
	note = {arXiv:1701.04862 [stat]},
}

@article{hess_deep_2023,
	title = {Deep {Learning} for {Bias}-{Correcting} {CMIP6}-{Class} {Earth} {System} {Models}},
	volume = {11},
	copyright = {© 2023 The Authors.},
	issn = {2328-4277},
	url = {https://onlinelibrary.wiley.com/doi/abs/10.1029/2023EF004002},
	doi = {10.1029/2023EF004002},
	language = {en},
	number = {10},
	urldate = {2025-06-16},
	journal = {Earth's Future},
	author = {Hess, Philipp and Lange, Stefan and Schötz, Christof and Boers, Niklas},
	year = {2023},
	note = {\_eprint: https://agupubs.onlinelibrary.wiley.com/doi/pdf/10.1029/2023EF004002},
	pages = {e2023EF004002},
}

@article{harris_generative_2022,
	title = {A {Generative} {Deep} {Learning} {Approach} to {Stochastic} {Downscaling} of {Precipitation} {Forecasts}},
	volume = {14},
	copyright = {© 2022 The Authors. Journal of Advances in Modeling Earth Systems published by Wiley Periodicals LLC on behalf of American Geophysical Union.},
	issn = {1942-2466},
	url = {https://onlinelibrary.wiley.com/doi/abs/10.1029/2022MS003120},
	doi = {10.1029/2022MS003120},
	language = {en},
	number = {10},
	urldate = {2025-06-16},
	journal = {Journal of Advances in Modeling Earth Systems},
	author = {Harris, Lucy and McRae, Andrew T. T. and Chantry, Matthew and Dueben, Peter D. and Palmer, Tim N.},
	year = {2022},
	note = {\_eprint: https://agupubs.onlinelibrary.wiley.com/doi/pdf/10.1029/2022MS003120},
	pages = {e2022MS003120},
}

@article{bochow_projections_2024,
	title = {Projections of precipitation and temperatures in {Greenland} and the impact of spatially uniform anomalies on the evolution of the ice sheet},
	volume = {18},
	issn = {1994-0416},
	url = {https://tc.copernicus.org/articles/18/5825/2024/},
	doi = {10.5194/tc-18-5825-2024},
	language = {English},
	number = {12},
	urldate = {2025-06-12},
	journal = {The Cryosphere},
	author = {Bochow, Nils and Poltronieri, Anna and Boers, Niklas},
	month = dec,
	year = {2024},
	note = {Publisher: Copernicus GmbH},
	pages = {5825--5863},
}

@article{glaude_factor_2024,
	title = {A {Factor} {Two} {Difference} in 21st-{Century} {Greenland} {Ice} {Sheet} {Surface} {Mass} {Balance} {Projections} {From} {Three} {Regional} {Climate} {Models} {Under} a {Strong} {Warming} {Scenario} ({SSP5}-8.5)},
	volume = {51},
	copyright = {© 2024. The Author(s).},
	issn = {1944-8007},
	url = {https://onlinelibrary.wiley.com/doi/abs/10.1029/2024GL111902},
	doi = {10.1029/2024GL111902},
	language = {en},
	number = {22},
	urldate = {2025-05-01},
	journal = {Geophysical Research Letters},
	author = {Glaude, Q. and Noel, B. and Olesen, M. and Van den Broeke, M. and van de Berg, W. J. and Mottram, R. and Hansen, N. and Delhasse, A. and Amory, C. and Kittel, C. and Goelzer, H. and Fettweis, X.},
	year = {2024},
	note = {\_eprint: https://onlinelibrary.wiley.com/doi/pdf/10.1029/2024GL111902},
	pages = {e2024GL111902},
}

@article{harder_hard-constrained_2023,
	title = {Hard-constrained deep learning for climate downscaling},
	volume = {24},
	issn = {1532-4435},
	number = {1},
	journal = {J. Mach. Learn. Res.},
	author = {Harder, Paula and Hernandez-Garcia, Alex and Ramesh, Venkatesh and Yang, Qidong and Sattegeri, Prasanna and Szwarcman, Daniela and Watson, Campbell D. and Rolnick, David},
	month = jul,
	year = {2023},
	pages = {365:17534--365:17573},
}

@misc{aich_diffusion_2025,
	title = {Diffusion models for probabilistic precipitation generation from atmospheric variables},
	url = {http://arxiv.org/abs/2504.00307},
	doi = {10.48550/arXiv.2504.00307},
	urldate = {2025-04-14},
	publisher = {arXiv},
	author = {Aich, Michael and Bathiany, Sebastian and Hess, Philipp and Huang, Yu and Boers, Niklas},
	month = apr,
	year = {2025},
	note = {arXiv:2504.00307 [cs]},
}

@misc{zhang_unreasonable_2018,
	title = {The {Unreasonable} {Effectiveness} of {Deep} {Features} as a {Perceptual} {Metric}},
	url = {http://arxiv.org/abs/1801.03924},
	doi = {10.48550/arXiv.1801.03924},
	urldate = {2025-04-08},
	publisher = {arXiv},
	author = {Zhang, Richard and Isola, Phillip and Efros, Alexei A. and Shechtman, Eli and Wang, Oliver},
	month = apr,
	year = {2018},
	note = {arXiv:1801.03924 [cs]},
}

@misc{song_consistency_2023,
	title = {Consistency {Models}},
	url = {http://arxiv.org/abs/2303.01469},
	doi = {10.48550/arXiv.2303.01469},
	urldate = {2025-04-08},
	publisher = {arXiv},
	author = {Song, Yang and Dhariwal, Prafulla and Chen, Mark and Sutskever, Ilya},
	month = may,
	year = {2023},
	note = {arXiv:2303.01469 [cs]},
}

@article{verjans_accelerating_2024,
	title = {Accelerating {Subglacial} {Hydrology} for {Ice} {Sheet} {Models} {With} {Deep} {Learning} {Methods}},
	volume = {51},
	copyright = {© 2024. The Authors.},
	issn = {1944-8007},
	url = {https://onlinelibrary.wiley.com/doi/abs/10.1029/2023GL105281},
	doi = {10.1029/2023GL105281},
	language = {en},
	number = {2},
	urldate = {2025-04-08},
	journal = {Geophysical Research Letters},
	author = {Verjans, Vincent and Robel, Alexander},
	year = {2024},
	note = {\_eprint: https://onlinelibrary.wiley.com/doi/pdf/10.1029/2023GL105281},
	pages = {e2023GL105281},
}

@article{steidl_physics-aware_2025,
	title = {Physics-aware machine learning for glacier ice thickness estimation: a case study for {Svalbard}},
	volume = {19},
	issn = {1994-0416},
	shorttitle = {Physics-aware machine learning for glacier ice thickness estimation},
	url = {https://tc.copernicus.org/articles/19/645/2025/},
	doi = {10.5194/tc-19-645-2025},
	language = {English},
	number = {2},
	urldate = {2025-04-08},
	journal = {The Cryosphere},
	author = {Steidl, Viola and Bamber, Jonathan Louis and Zhu, Xiao Xiang},
	month = feb,
	year = {2025},
	note = {Publisher: Copernicus GmbH},
	pages = {645--661},
}

@article{van_der_meer_minimal_2025,
	title = {A minimal machine-learning glacier mass balance model},
	volume = {19},
	issn = {1994-0416},
	url = {https://tc.copernicus.org/articles/19/805/2025/},
	doi = {10.5194/tc-19-805-2025},
	language = {English},
	number = {2},
	urldate = {2025-04-08},
	journal = {The Cryosphere},
	author = {van der Meer, Marijn and Zekollari, Harry and Huss, Matthias and Bolibar, Jordi and Sjursen, Kamilla Hauknes and Farinotti, Daniel},
	month = feb,
	year = {2025},
	note = {Publisher: Copernicus GmbH},
	pages = {805--826},
}

@article{bolibar_deep_2020,
	title = {Deep learning applied to glacier evolution modelling},
	volume = {14},
	issn = {1994-0416},
	url = {https://tc.copernicus.org/articles/14/565/2020/},
	doi = {10.5194/tc-14-565-2020},
	language = {English},
	number = {2},
	urldate = {2025-04-08},
	journal = {The Cryosphere},
	author = {Bolibar, Jordi and Rabatel, Antoine and Gouttevin, Isabelle and Galiez, Clovis and Condom, Thomas and Sauquet, Eric},
	month = feb,
	year = {2020},
	note = {Publisher: Copernicus GmbH},
	pages = {565--584},
}

@article{rosier_predicting_2023,
	title = {Predicting ocean-induced ice-shelf melt rates using deep learning},
	volume = {17},
	issn = {1994-0416},
	url = {https://tc.copernicus.org/articles/17/499/2023/},
	doi = {10.5194/tc-17-499-2023},
	language = {English},
	number = {2},
	urldate = {2025-04-08},
	journal = {The Cryosphere},
	author = {Rosier, Sebastian H. R. and Bull, Christopher Y. S. and Woo, Wai L. and Gudmundsson, G. Hilmar},
	month = feb,
	year = {2023},
	note = {Publisher: Copernicus GmbH},
	pages = {499--518},
}

@article{wang_deep_2025,
	title = {Deep learning the flow law of {Antarctic} ice shelves},
	volume = {387},
	url = {https://www.science.org/doi/10.1126/science.adp3300},
	doi = {10.1126/science.adp3300},
	number = {6739},
	urldate = {2025-04-08},
	journal = {Science},
	author = {Wang, Yongji and Lai, Ching-Yao and Prior, David J. and Cowen-Breen, Charlie},
	month = mar,
	year = {2025},
	note = {Publisher: American Association for the Advancement of Science},
	pages = {1219--1224},
}

@article{jouvet_deep_2022,
	title = {Deep learning speeds up ice flow modelling by several orders of magnitude},
	volume = {68},
	issn = {0022-1430, 1727-5652},
	url = {https://www.cambridge.org/core/journals/journal-of-glaciology/article/deep-learning-speeds-up-ice-flow-modelling-by-several-orders-of-magnitude/748E962A103D2AF45F4CA8823C88C0C0},
	doi = {10.1017/jog.2021.120},
	language = {en},
	number = {270},
	urldate = {2025-04-08},
	journal = {Journal of Glaciology},
	author = {Jouvet, Guillaume and Cordonnier, Guillaume and Kim, Byungsoo and Lüthi, Martin and Vieli, Andreas and Aschwanden, Andy},
	month = aug,
	year = {2022},
	pages = {651--664},
}

@article{jouvet_ice-flow_2023,
	title = {Ice-flow model emulator based on physics-informed deep learning},
	volume = {69},
	issn = {0022-1430, 1727-5652},
	url = {https://www.cambridge.org/core/journals/journal-of-glaciology/article/iceflow-model-emulator-based-on-physicsinformed-deep-learning/8C4D103C0F34DA690D9B524DF1461C5C},
	doi = {10.1017/jog.2023.73},
	language = {en},
	number = {278},
	urldate = {2025-04-08},
	journal = {Journal of Glaciology},
	author = {Jouvet, Guillaume and Cordonnier, Guillaume},
	month = dec,
	year = {2023},
	pages = {1941--1955},
}

@article{lam_learning_2023,
	title = {Learning skillful medium-range global weather forecasting},
	volume = {382},
	url = {https://www.science.org/doi/10.1126/science.adi2336},
	doi = {10.1126/science.adi2336},
	number = {6677},
	urldate = {2025-04-08},
	journal = {Science},
	author = {Lam, Remi and Sanchez-Gonzalez, Alvaro and Willson, Matthew and Wirnsberger, Peter and Fortunato, Meire and Alet, Ferran and Ravuri, Suman and Ewalds, Timo and Eaton-Rosen, Zach and Hu, Weihua and Merose, Alexander and Hoyer, Stephan and Holland, George and Vinyals, Oriol and Stott, Jacklynn and Pritzel, Alexander and Mohamed, Shakir and Battaglia, Peter},
	month = dec,
	year = {2023},
	note = {Publisher: American Association for the Advancement of Science},
	pages = {1416--1421},
}

@article{kochkov_neural_2024,
	title = {Neural general circulation models for weather and climate},
	volume = {632},
	copyright = {2024 The Author(s)},
	issn = {1476-4687},
	url = {https://www.nature.com/articles/s41586-024-07744-y},
	doi = {10.1038/s41586-024-07744-y},
	language = {en},
	number = {8027},
	urldate = {2025-04-08},
	journal = {Nature},
	author = {Kochkov, Dmitrii and Yuval, Janni and Langmore, Ian and Norgaard, Peter and Smith, Jamie and Mooers, Griffin and Klöwer, Milan and Lottes, James and Rasp, Stephan and Düben, Peter and Hatfield, Sam and Battaglia, Peter and Sanchez-Gonzalez, Alvaro and Willson, Matthew and Brenner, Michael P. and Hoyer, Stephan},
	month = aug,
	year = {2024},
	note = {Publisher: Nature Publishing Group},
	pages = {1060--1066},
}

@article{bochow_reconstructing_2025,
	title = {Reconstructing historical climate fields with deep learning},
	volume = {11},
	url = {https://www.science.org/doi/10.1126/sciadv.adp0558},
	doi = {10.1126/sciadv.adp0558},
	number = {14},
	urldate = {2025-04-08},
	journal = {Science Advances},
	author = {Bochow, Nils and Poltronieri, Anna and Rypdal, Martin and Boers, Niklas},
	month = apr,
	year = {2025},
	note = {Publisher: American Association for the Advancement of Science},
	pages = {eadp0558},
}

@article{hess_fast_2025,
	title = {Fast, scale-adaptive and uncertainty-aware downscaling of {Earth} system model fields with generative machine learning},
	volume = {7},
	copyright = {2025 The Author(s)},
	issn = {2522-5839},
	url = {https://www.nature.com/articles/s42256-025-00980-5},
	doi = {10.1038/s42256-025-00980-5},
	language = {en},
	number = {3},
	urldate = {2025-04-07},
	journal = {Nature Machine Intelligence},
	author = {Hess, Philipp and Aich, Michael and Pan, Baoxiang and Boers, Niklas},
	month = mar,
	year = {2025},
	note = {Publisher: Nature Publishing Group},
	pages = {363--373},
}

@article{seroussi_ismip6_2020,
	title = {{ISMIP6} {Antarctica}: a multi-model ensemble of the {Antarctic} ice sheet evolution over the 21st century},
	volume = {14},
	issn = {1994-0416},
	shorttitle = {{ISMIP6} {Antarctica}},
	url = {https://tc.copernicus.org/articles/14/3033/2020/},
	doi = {10.5194/tc-14-3033-2020},
	language = {English},
	number = {9},
	urldate = {2025-03-04},
	journal = {The Cryosphere},
	author = {Seroussi, Hélène and Nowicki, Sophie and Payne, Antony J. and Goelzer, Heiko and Lipscomb, William H. and Abe-Ouchi, Ayako and Agosta, Cécile and Albrecht, Torsten and Asay-Davis, Xylar and Barthel, Alice and Calov, Reinhard and Cullather, Richard and Dumas, Christophe and Galton-Fenzi, Benjamin K. and Gladstone, Rupert and Golledge, Nicholas R. and Gregory, Jonathan M. and Greve, Ralf and Hattermann, Tore and Hoffman, Matthew J. and Humbert, Angelika and Huybrechts, Philippe and Jourdain, Nicolas C. and Kleiner, Thomas and Larour, Eric and Leguy, Gunter R. and Lowry, Daniel P. and Little, Chistopher M. and Morlighem, Mathieu and Pattyn, Frank and Pelle, Tyler and Price, Stephen F. and Quiquet, Aurélien and Reese, Ronja and Schlegel, Nicole-Jeanne and Shepherd, Andrew and Simon, Erika and Smith, Robin S. and Straneo, Fiammetta and Sun, Sainan and Trusel, Luke D. and Van Breedam, Jonas and van de Wal, Roderik S. W. and Winkelmann, Ricarda and Zhao, Chen and Zhang, Tong and Zwinger, Thomas},
	month = sep,
	year = {2020},
	note = {Publisher: Copernicus GmbH},
	pages = {3033--3070},
}

@article{nowicki_experimental_2020,
	title = {Experimental protocol for sea level projections from {ISMIP6} stand-alone ice sheet models},
	volume = {14},
	issn = {1994-0416},
	url = {https://tc.copernicus.org/articles/14/2331/2020/},
	doi = {10.5194/tc-14-2331-2020},
	language = {English},
	number = {7},
	urldate = {2025-03-04},
	journal = {The Cryosphere},
	author = {Nowicki, Sophie and Goelzer, Heiko and Seroussi, Hélène and Payne, Anthony J. and Lipscomb, William H. and Abe-Ouchi, Ayako and Agosta, Cécile and Alexander, Patrick and Asay-Davis, Xylar S. and Barthel, Alice and Bracegirdle, Thomas J. and Cullather, Richard and Felikson, Denis and Fettweis, Xavier and Gregory, Jonathan M. and Hattermann, Tore and Jourdain, Nicolas C. and Kuipers Munneke, Peter and Larour, Eric and Little, Christopher M. and Morlighem, Mathieu and Nias, Isabel and Shepherd, Andrew and Simon, Erika and Slater, Donald and Smith, Robin S. and Straneo, Fiammetta and Trusel, Luke D. and van den Broeke, Michiel R. and van de Wal, Roderik},
	month = jul,
	year = {2020},
	note = {Publisher: Copernicus GmbH},
	pages = {2331--2368},
}

@article{beckmann_effects_2023,
	title = {Effects of extreme melt events on ice flow and sea level rise of the {Greenland} {Ice} {Sheet}},
	volume = {17},
	issn = {1994-0416},
	url = {https://tc.copernicus.org/articles/17/3083/2023/},
	doi = {10.5194/tc-17-3083-2023},
	language = {English},
	number = {7},
	urldate = {2025-03-04},
	journal = {The Cryosphere},
	author = {Beckmann, Johanna and Winkelmann, Ricarda},
	month = jul,
	year = {2023},
	note = {Publisher: Copernicus GmbH},
	pages = {3083--3099},
}

@article{zeitz_impact_2021,
	title = {Impact of the melt–albedo feedback on the future evolution of the {Greenland} {Ice} {Sheet} with {PISM}-{dEBM}-simple},
	volume = {15},
	issn = {1994-0416},
	url = {https://tc.copernicus.org/articles/15/5739/2021/},
	doi = {10.5194/tc-15-5739-2021},
	language = {English},
	number = {12},
	urldate = {2025-03-04},
	journal = {The Cryosphere},
	author = {Zeitz, Maria and Reese, Ronja and Beckmann, Johanna and Krebs-Kanzow, Uta and Winkelmann, Ricarda},
	month = dec,
	year = {2021},
	note = {Publisher: Copernicus GmbH},
	pages = {5739--5764},
}

@article{bochow_overshooting_2023,
	title = {Overshooting the critical threshold for the {Greenland} ice sheet},
	volume = {622},
	copyright = {2023 The Author(s)},
	issn = {1476-4687},
	url = {https://www.nature.com/articles/s41586-023-06503-9},
	doi = {10.1038/s41586-023-06503-9},
	language = {en},
	number = {7983},
	urldate = {2025-03-04},
	journal = {Nature},
	author = {Bochow, Nils and Poltronieri, Anna and Robinson, Alexander and Montoya, Marisa and Rypdal, Martin and Boers, Niklas},
	month = oct,
	year = {2023},
	note = {Publisher: Nature Publishing Group},
	pages = {528--536},
}

@article{aschwanden_contribution_2019,
	title = {Contribution of the {Greenland} {Ice} {Sheet} to sea level over the next millennium},
	volume = {5},
	url = {https://www.science.org/doi/10.1126/sciadv.aav9396},
	doi = {10.1126/sciadv.aav9396},
	number = {6},
	urldate = {2025-03-04},
	journal = {Science Advances},
	author = {Aschwanden, Andy and Fahnestock, Mark A. and Truffer, Martin and Brinkerhoff, Douglas J. and Hock, Regine and Khroulev, Constantine and Mottram, Ruth and Khan, S. Abbas},
	month = jun,
	year = {2019},
	note = {Publisher: American Association for the Advancement of Science},
	pages = {eaav9396},
}

@article{garbe_hysteresis_2020,
	title = {The hysteresis of the {Antarctic} {Ice} {Sheet}},
	volume = {585},
	copyright = {2020 The Author(s), under exclusive licence to Springer Nature Limited},
	issn = {1476-4687},
	url = {https://www.nature.com/articles/s41586-020-2727-5},
	doi = {10.1038/s41586-020-2727-5},
	language = {en},
	number = {7826},
	urldate = {2025-03-04},
	journal = {Nature},
	author = {Garbe, Julius and Albrecht, Torsten and Levermann, Anders and Donges, Jonathan F. and Winkelmann, Ricarda},
	month = sep,
	year = {2020},
	note = {Publisher: Nature Publishing Group},
	pages = {538--544},
}

@article{choi_ice_2021,
	title = {Ice dynamics will remain a primary driver of {Greenland} ice sheet mass loss over the next century},
	volume = {2},
	copyright = {2021 The Author(s)},
	issn = {2662-4435},
	url = {https://www.nature.com/articles/s43247-021-00092-z},
	doi = {10.1038/s43247-021-00092-z},
	language = {en},
	number = {1},
	urldate = {2025-03-04},
	journal = {Communications Earth \& Environment},
	author = {Choi, Youngmin and Morlighem, Mathieu and Rignot, Eric and Wood, Michael},
	month = feb,
	year = {2021},
	note = {Publisher: Nature Publishing Group},
	pages = {1--9},
}

@article{cannon_bias_2015,
	title = {Bias {Correction} of {GCM} {Precipitation} by {Quantile} {Mapping}: {How} {Well} {Do} {Methods} {Preserve} {Changes} in {Quantiles} and {Extremes}?},
	volume = {28},
	issn = {0894-8755, 1520-0442},
	shorttitle = {Bias {Correction} of {GCM} {Precipitation} by {Quantile} {Mapping}},
	url = {https://journals.ametsoc.org/view/journals/clim/28/17/jcli-d-14-00754.1.xml},
	doi = {10.1175/JCLI-D-14-00754.1},
	language = {EN},
	number = {17},
	urldate = {2025-03-04},
	journal = {Journal of Climate},
	author = {Cannon, Alex J. and Sobie, Stephen R. and Murdock, Trevor Q.},
	month = sep,
	year = {2015},
	note = {Publisher: American Meteorological Society
Section: Journal of Climate},
	pages = {6938--6959},
}

@article{fettweis_reconstructions_2017,
	title = {Reconstructions of the 1900–2015 {Greenland} ice sheet surface mass balance using the regional climate {MAR} model},
	volume = {11},
	issn = {1994-0416},
	url = {https://tc.copernicus.org/articles/11/1015/2017/},
	doi = {10.5194/tc-11-1015-2017},
	language = {English},
	number = {2},
	urldate = {2025-02-18},
	journal = {The Cryosphere},
	author = {Fettweis, Xavier and Box, Jason E. and Agosta, Cécile and Amory, Charles and Kittel, Christoph and Lang, Charlotte and van As, Dirk and Machguth, Horst and Gallée, Hubert},
	month = apr,
	year = {2017},
	note = {Publisher: Copernicus GmbH},
	pages = {1015--1033},
}

@article{fettweis_estimating_2013,
	title = {Estimating the {Greenland} ice sheet surface mass balance contribution to future sea level rise using the regional atmospheric climate model {MAR}},
	volume = {7},
	issn = {1994-0416},
	url = {https://tc.copernicus.org/articles/7/469/2013/},
	doi = {10.5194/tc-7-469-2013},
	language = {English},
	number = {2},
	urldate = {2025-02-18},
	journal = {The Cryosphere},
	author = {Fettweis, X. and Franco, B. and Tedesco, M. and van Angelen, J. H. and Lenaerts, J. T. M. and van den Broeke, M. R. and Gallée, H.},
	month = mar,
	year = {2013},
	note = {Publisher: Copernicus GmbH},
	pages = {469--489},
}

\end{document}